\documentclass[aps,prl,reprint,preprintnumbers,superscriptaddress,amsmath,amssymb,bibnotes,longbibliography]{revtex4-2}
\usepackage{graphicx}
\usepackage{dcolumn}
\usepackage{bm}
\usepackage[colorlinks,linkcolor=blue,anchorcolor=blue,citecolor=blue,breaklinks,CJKbookmarks=True,urlcolor=blue,filecolor=blue,menucolor=blue,runcolor=blue]{hyperref}

\begin{document}

\title{Fully gapped superconductivity with preserved time reversal symmetry in noncentrosymmetric LaPdIn }

\author{H. Su}
\affiliation  {Center for Correlated Matter and Department of Physics, Zhejiang University, Hangzhou 310058, China}
\affiliation  {Zhejiang Province Key Laboratory of Quantum Technology and Device, Department of Physics, Zhejiang University, Hangzhou 310058, China}
\author{Z. Y. Nie}
\affiliation  {Center for Correlated Matter and Department of Physics, Zhejiang University, Hangzhou 310058, China}
\affiliation  {Zhejiang Province Key Laboratory of Quantum Technology and Device, Department of Physics, Zhejiang University, Hangzhou 310058, China}
\author{F. Du}
\affiliation  {Center for Correlated Matter and Department of Physics, Zhejiang University, Hangzhou 310058, China}
\affiliation  {Zhejiang Province Key Laboratory of Quantum Technology and Device, Department of Physics, Zhejiang University, Hangzhou 310058, China}
\author{S. S. Luo}
\affiliation  {Center for Correlated Matter and Department of Physics, Zhejiang University, Hangzhou 310058, China}
\affiliation  {Zhejiang Province Key Laboratory of Quantum Technology and Device, Department of Physics, Zhejiang University, Hangzhou 310058, China}
\author{A. Wang}
\affiliation  {Center for Correlated Matter and Department of Physics, Zhejiang University, Hangzhou 310058, China}
\affiliation  {Zhejiang Province Key Laboratory of Quantum Technology and Device, Department of Physics, Zhejiang University, Hangzhou 310058, China}
\author{Y. J. Zhang}
\affiliation  {Center for Correlated Matter and Department of Physics, Zhejiang University, Hangzhou 310058, China}
\affiliation  {Institute for Advanced Materials, Hubei Normal University, Huangshi 435002, China}
\author{Y. Chen}
\affiliation  {Center for Correlated Matter and Department of Physics, Zhejiang University, Hangzhou 310058, China}
\affiliation  {Zhejiang Province Key Laboratory of Quantum Technology and Device, Department of Physics, Zhejiang University, Hangzhou 310058, China}
\author{P. K. Biswas}
\affiliation  {ISIS Facility, STFC Rutherford Appleton Laboratory, Harwell Science and Innovation Campus, Oxfordshire, OX11 0QX, United Kingdom}
\author{D. T. Adroja}
\affiliation  {ISIS Facility, STFC Rutherford Appleton Laboratory, Harwell Science and Innovation Campus, Oxfordshire, OX11 0QX, United Kingdom}
\affiliation  {Highly Correlated Matter Research Group, Physics Department, University of Johannesburg, P.O. Box 524,
Auckland Park 2006, South Africa}
\author{C. Cao}
\affiliation  {Center for Correlated Matter and Department of Physics, Zhejiang University, Hangzhou 310058, China}
\affiliation  {Condensed Matter Group, Department of Physics, Hangzhou Normal University, Hangzhou 311121, China}
\author{M. Smidman}
\email[Corresponding author: ]{msmidman@zju.edu.cn}
\affiliation  {Center for Correlated Matter and Department of Physics, Zhejiang University, Hangzhou 310058, China}
\affiliation  {Zhejiang Province Key Laboratory of Quantum Technology and Device, Department of Physics, Zhejiang University, Hangzhou 310058, China}
\author{H. Q. Yuan}
\email[Corresponding author: ]{hqyuan@zju.edu.cn}
\affiliation  {Center for Correlated Matter and Department of Physics, Zhejiang University, Hangzhou 310058, China}
\affiliation  {Zhejiang Province Key Laboratory of Quantum Technology and Device, Department of Physics, Zhejiang University, Hangzhou 310058, China}
\affiliation  {State Key Laboratory of Silicon Materials, Zhejiang University, Hangzhou 310058, China}
\affiliation  {Collaborative Innovation Center of Advanced Microstructures, Nanjing University, Nanjing, 210093, China}

\date{\today}


\begin{abstract}
We report an investigation of the superconducting properties of the hexagonal noncentrosymmetric compound LaPdIn. Electrical resistivity, specific heat and ac susceptibility measurements demonstrate the presence of bulk superconductivity below $T_c$ = 1.6~K. The specific heat, together with the penetration depth measured using transverse-field muon spin rotation and the tunnel diode oscillator based method, are well described by single gap $s$-wave superconductivity, with a gap magnitude of 1.8$k_BT_c$. From zero-field muon spin relaxation results no evidence is found for the spontaneous emergence of magnetic fields in the superconducting state, indicating that time-reversal symmetry is preserved. Band structure calculations reveal that there is a relatively weak effect of antisymmetric spin-orbit coupling on the electronic bands near the Fermi level, which is consistent with there being negligible singlet-triplet mixing due to broken inversion symmetry. On the other hand, isostructural LuPdIn and LaPtIn do not exhibit superconductivity down to 0.4~K, which may be due to these systems having a smaller density of states at the Fermi level.

\end{abstract}

\maketitle


\section{\uppercase\expandafter{\romannumeral1}. INTRODUCTION}

Since the discovery of superconductivity in the noncentrosymmetric heavy fermion compound CePt$_3$Si \cite{Bauer2004}, there has been considerable interest in superconductors with broken inversion symmetry in the crystal structure, due to the potential for unusual superconducting properties \cite{bauer2012, Smidman2017}. In a noncentrosymmetric material, the lack of an inversion center leads to antisymmetric spin-orbit coupling (ASOC), lifting the electron spin degeneracy. As parity is no longer a good quantum number, the ASOC can give rise to Cooper pairs which are in a mixture of spin singlet and triplet states. Under such circumstances, unusual properties including nodal superconducting gaps, anisotropic upper critical fields exceeding the Pauli limit along some directions, and Majorana zero modes \cite{SatoPhysRevB2009, SatoPhysRevB2010} are anticipated.

Besides CePt$_3$Si, non-centrosymmetric heavy-fermion superconductiviy is also found in CeTX$_3$ (T = transition metal, X = Si/Ge) \cite{Kimura2005, Sugitani2006, Settai2007, Honda2010} and UIr \cite{Akazawa2004}. In such f-electron superconductors, unconventional pairing states are generally expected due to the large Coulomb repulsion and presence of spin fluctuations. As such, it can be difficult to disentangle the role of these factors from that of broken inversion symmetry. It has therefore been of interest to also examine non-magnetic weakly correlated noncentrosymmetric superconductors (NCS). In the case of Li$_2$(Pd$_{1-x}$Pt$_x$)$_3$B, increasing the Pt concentration tunes the strength of the ASOC, leading to a change from a predominantly spin singlet state with two nodeless gaps in Li$_2$Pd$_3$B, to a triplet dominant state in Li$_2$Pt$_3$B with line nodes in one of the gaps \cite{Yuan2006}. Evidence for nodes in the superconducting gap is also found for Y$_2$C$_3$ \cite{Chen2011}, K$_2$Cr$_3$As$_3$ \cite{Adrojaprb2015, Panggm2015} and CaPtAs \cite{Xie2020, Shangprl2020}. Meanwhile, K$_2$Cr$_3$As$_3$ exhibits extremely large and anisotropic upper critical fields, evidencing the absence of Pauli limiting along some field directions \cite{BaoPhysRevX2015, KongPhysRevB2015}.

Signatures of time-reversal symmetry breaking (TRSB), where spontaneous magnetic fields emerge in the superconducting state, have also been detected in some noncentrosymmetric superconductors, including LaNiC$_2$ \cite{Hillier}, La$_7$(Ir,Rh)$_3$ \cite{Barker2015, SinghPhysRevB2020}, Re-(Zr,Hf,Ti,Nb) alloys \cite{Singh2014, Shangprl2018, SinghPhysRevB2017, SinghPhysRevB2018} and CaPtAs \cite{Shangprl2020}. However, the connection between broken inversion symmetry in the crystal structure and broken time reversal symmetry remains unclear. On the other hand, the properties of many NCS are consistent with conventional single gap $s$-wave superconductivity, where the presence of ASOC does not appear to be sufficient to lead to unconventional pairing states \cite{Smidman2017}. It is therefore important to examine and characterize the pairing states of different families of noncentrosymmetric superconductors.

A large number of intermetallic compounds crystallize in the ZrNiAl-type structure (space group $P\bar{6}2m$) \cite{Markus1999}, where the Zr atoms form a Kagome lattice within the basal plane, and there are alternate layers of Zr-Ni and Ni-Al  stacked along the $c$-axis. Among this family, (Zr,Hf)RuP \cite{Barz1980}, ZrRu(As,Si) \cite{MEISNER1983, Ichimin1999, das2021probing} and Mo(Ni,Ru)P \cite{SHIROTANI2000} are superconductors with a relatively high $T_c$  over 10~K. Meanwhile the rare earth based compounds ScIrP, ScRhP, LaRhSn and LaPdIn have lower $T_c$ values \cite{Okamoto2016, Inohara2016, MIHALIK2008, Bruck1988}. Detailed characterizations of the order parameters of this family of superconductors are required to both determine whether time reversal symmetry is preserved in the superconducting state, and to look for the presence of singlet-triplet mixing or other effects arising from the ASOC. Here we probe the superconducting order parameter of LaPdIn, by measuring the magnetic penetration depth utilizing the muon-spin relaxation/rotation ($\mu$SR) and the tunnel diode oscillator based methods, as well as the specific heat. We find that time-reversal symmetry is preserved within the superconducting state, and the results are well accounted for by single-gap $s$-wave superconductivity.

\section{\uppercase\expandafter{\romannumeral2}. EXPERIMENTAL METHODS}

Polycrystalline LaPdIn was synthesized via arc-melting stoichiometric quantities of the starting materials La, Pd and In under an argon atmosphere. The sample was annealed at 700~$^{\circ}$C for one week in an evacuated quartz tube. In addition, polycrystalline samples of LuPdIn and LaPtIn were synthesized using the same method. The crystal structure and phase purity were checked at room temperature using powder x-ray diffraction with Cu-K$\alpha$ radiation. The electrical resistivity down to 0.4~K was measured using a standard four probe method in a Quantum Design Physical Property Measurement System (PPMS) Dynacool equipped with a $^3$He insert, using a 1~mA 15~Hz excitation current. Heat capacity measurements down to 0.4~K were also performed in the PPMS with a $^3$He insert, using the relaxation method. Alternating-current (ac) magnetic susceptibility measurements were carried out in an Oxford Instruments $^3$He refrigerator equipped with a commercial magnetometer with a frequency of the excitation field of 633~Hz.

Muon spin relaxation/rotation ($\mu$SR) measurements were carried out on the MuSR spectrometer at the ISIS facility of the Rutherford Appleton Laboratory, UK. The sample was powdered, mounted on a silver plate, and cooled in a dilution refrigerator. $\mu$SR measurements were performed in both zero-field, where the magnetic field was canceled using an active compensation system to within 1~$\mu$T, as well as in various transverse fields, applied perpendicular to the initial polarization direction of the muons.

The temperature dependence of the London penetration depth was also probed using the tunnel-diode oscillator (TDO) based technique carried out in a $^3$He cryostat \cite{Degrift1975}. Cuboid samples with dimensions less than 1$\times$1$\times$1~mm$^3$ were inserted into the TDO coil, with an excitation ac field of 20~mOe, and an operating frequency of around 7~MHz. The change of London penetration depth $\Delta\lambda(T)=\lambda(T)-\lambda(0)$ is related to the frequency shift $\Delta f(T)=f(T)-f(0)$ via $\Delta\lambda(T)=G\Delta f(T)$, where $G$ is a calibration factor related to the coil and sample geometry \cite{Prozorov2000}.

Calculations of the band structure and electronic density of states were performed using density functional theory (DFT), as implemented in the Vienna ab-initio simulation package, with the Perdew-Burke-Ernzerhoff (PBE) functional in the generalized gradient approximation (GGA). The Brillouin zone was sampled with $6\times6\times12$ $K$ mesh and the energy-cutoff was selected to be 450~eV to ensure convergence.

\section{\uppercase\expandafter{\romannumeral3}. Results}

\subsection{\uppercase\expandafter{a}. Crystal structure}
\begin{figure}
  	\includegraphics[angle=0,width=0.49\textwidth]{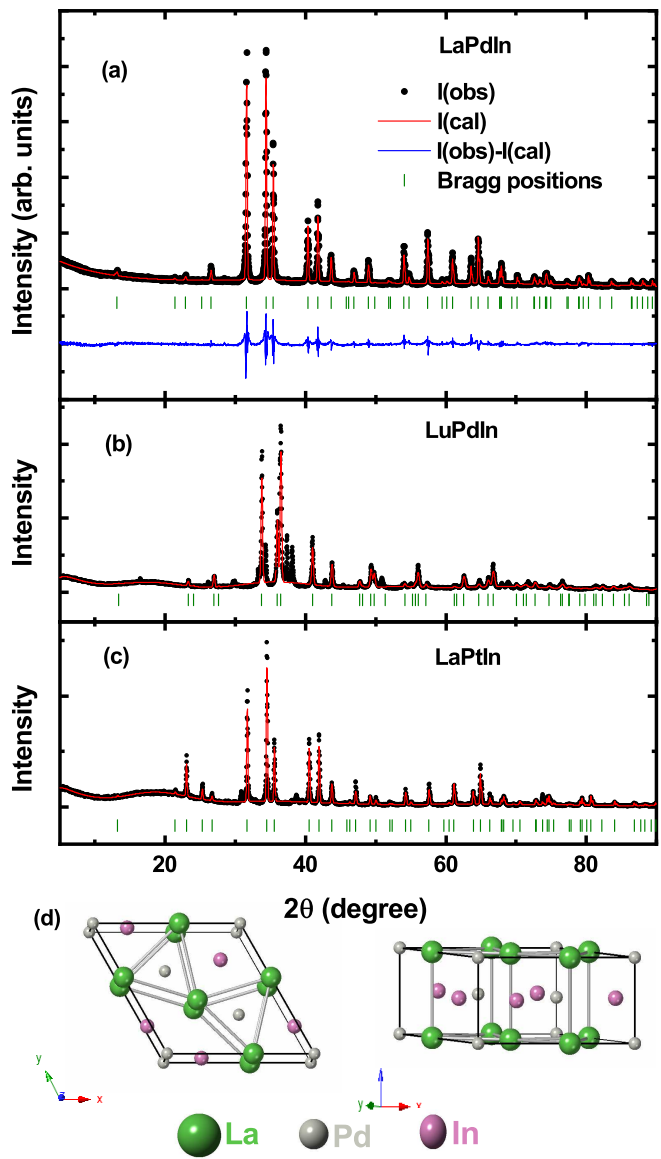}
  	\vspace{-12pt} \caption{\label{Figure1}(Color online)   Room temperature powder x-ray diffraction patterns of isostructural (a) LaPdIn, (b) LuPdIn and (c) LaPtIn. The red solid lines are calculated from the results of the Rietveld refinement based on the ZrNiAl-type structure, and the vertical bars mark the Bragg peak positions. The difference between the experimental data and calculated curve for LaPdIn are shown by the blue curve. (d) Crystal structure of LaPdIn, where the green, grey and pink balls represent La, Pd and In atoms respectively.  }
  	\vspace{-12pt}
\end{figure}

\begin{table}[tb]
\caption{Results of the Rietveld refinements of the powder x-ray diffraction measurements.}
\label{table:table1}
\begin{ruledtabular}
 \begin{tabular}{c c c c }
{} &LaPdIn  &LuPdIn  &LaPtIn  \\
\hline\\[-2ex]
{Crystal structure} &\multicolumn{3}{c}{ZrNiAl-type} \\
{Space group} &\multicolumn{3}{c}{P$\overline{6}$2m, No.189} \\
$a$(\AA)  &7.7384(2)  &7.6195(2)  &7.6994(1)  \\
$c$(\AA)  &4.1440(1)  &3.6915(1)  &4.1330(1)  \\
$c/a$     &0.5355  &0.4844  &0.5367  \\
$V_{\rm cell}$(\AA$^3$)    &214.91  &185.60     &212.18                          \\
$R_p$($\%$)  &10.35   &15.93   &10.77   \\
$R_{wp}$($\%$) &14.01  &23.65  &15.44   \\
\end{tabular}
\end{ruledtabular}
\end{table}

Figures \ref{Figure1}(a)-(c) display the powder x-ray diffraction patterns for LaPdIn,  LuPdIn and LaPtIn, where the diffraction peaks are indexed by the ZrNiAl-type structure, shown in Fig. \ref{Figure1}(d). The $c$-axis of the unit cell is much shorter than the $a$- and $b$- axes, where the La and In occupy noncentrosymmetric sites, while the Pd sites are centrosymmetric. In the case of LaPdIn, any peaks associated with impurity phases are very small, with an intensity less than $0.5\%$ of the largest sample peaks, while for LaPtIn and LuPdIn there are larger peaks from an unknown impurity phase. The results of the Rietveld refinements of the diffraction data are shown in Table. \ref{table:table1}, which are consistent with previous reports \cite{GONDEK2007}.

\subsection{\uppercase\expandafter{b}. Zero-field properties of LaPdIn}

  \begin{figure}
  	\includegraphics[angle=0,width=0.49\textwidth]{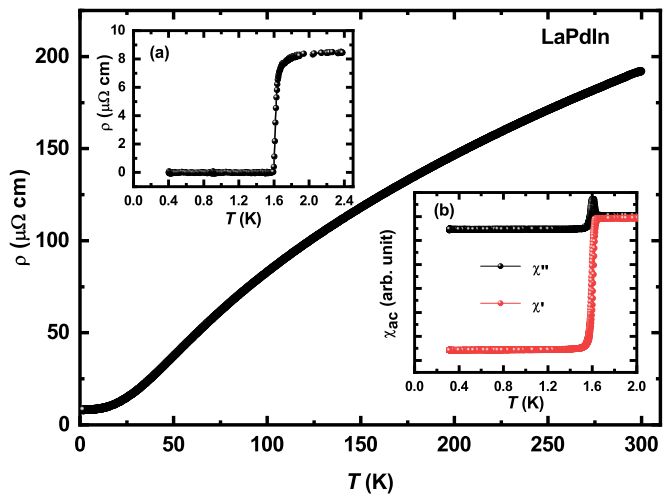}
  	\vspace{-12pt} \caption{\label{Figure2}(Color online)  Temperature dependence of the resistivity $\rho(T)$ of LaPdIn measured from 300~K to 0.4~K in zero field. Inset (a): low-temperature resistivity near the superconducting transition at $T_c$=1.6~K. Inset (b): the ac susceptibility in the vicinity of $T_c$.}
  	\vspace{-12pt}
  \end{figure}

  \begin{figure}
  	\includegraphics[angle=0,width=0.49\textwidth]{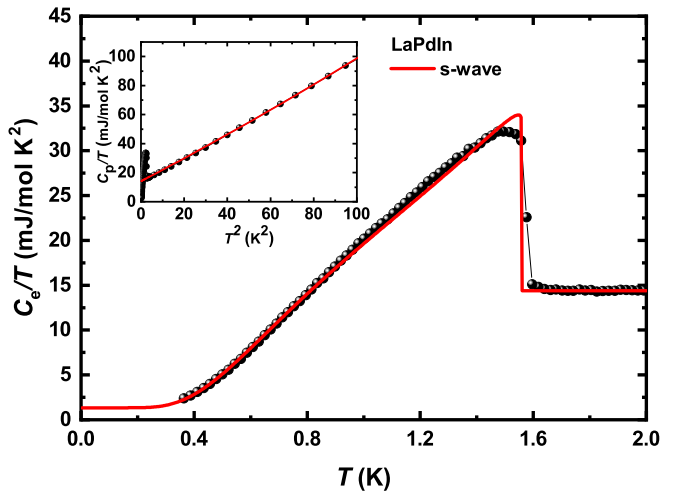}
  	\vspace{-12pt} \caption{\label{Figure3}(Color online) Electronic specific heat as $C_e/T$ of LaPdIn in zero field. The red solid line shows the fit to a single gap $s$-wave model with $\Delta(0)=1.80~k_B$T$_c$. Inset: Low temperature $C_p/T$ vs $T^2$, where the red solid line shows the result from fitting with $C_p/T=\gamma_n+\beta T^2+\delta T^4$. }
  	\vspace{-12pt}
  \end{figure}

The main panel of Figure \ref{Figure2} displays the temperature dependence of the resistivity of LaPdIn in zero field, which shows metallic behavior. The relatively low residual resistivity ($\rho_0$ = 8.32~$\mu\Omega$~cm) and moderate residual resistivity ratio (RRR = $\rho$(300~K)/$\rho$(1.6~K) = 23.6) indicate a good sample quality. The low temperature resistivity is displayed in the inset (a), where the resistivity value abruptly drops to zero at a superconducting transition with $T_c$ = 1.6~K. The inset (b) shows the low temperature ac susceptibility $\chi_{ac}$. The superconducting transition is inferred from a sharp drop of the real part $\chi'$, as well as a peak in the imaginary part $\chi''$ at $T_c$. The low temperature specific heat in the normal state (inset of Fig. \ref{Figure3}) is well described by $C_p/T$ = $\gamma_n+\beta T^2+\delta T^4$, $\beta$ and $\delta$ are the coefficients corresponding to the phonon contribution, with $\gamma_n$ = 14.46(6)~mJ~mol$^{-1}$~K$^{-2}$, $\beta$ = 0.777(5)~mJ~mol$^{-1}$~K$^{-4}$ and $\delta$ = 0.6887(1)~$\mu$J~mol$^{-1}$~K$^{-6}$. The relatively low $\gamma$ value suggests a small electronic effective-mass and weak electronic correlations. The Debye temperature $\theta_D$ is estimated to be 195(2)~K using $\theta_D=(12\pi^4Rn/5\beta)^{1/3}$, where $n$ = 3 is the number of atoms per formula unit and $R$ = 8.314~J~mol$^{-1}$~K$^{-1}$ is the molar gas constant. The electron-phonon coupling constant $\lambda_{el-ph}$ can be estimated using McMillan's theory, which relates $T_c$ and $\theta_D$ via
\begin{equation}\label{equation4}
\lambda_{el-ph}=\frac{1.04+\mu^*\mathrm{ln}(\frac{\theta_D}{1.45T_c})}{(1-0.62\mu^*)\mathrm{ln}(\frac{\theta_D}{1.45T_c})-1.04}
\end{equation}
For LaPdIn, using the typical range of values for $\mu^*$ of $0.1-0.15$, $\lambda_{el-ph}=0.48-0.57$ are obtained, corresponding to weakly coupled superconductivity.

After subtracting the phonon contribution from the total specific heat, the temperature dependence of the electronic contribution $C_e/T$ is shown in Fig. \ref{Figure3}. The normalized specific heat jump at $T_c$, $\Delta C_{e}/\gamma_nT_c$ is 1.23, which corresponds to bulk superconductivity and is slightly smaller than the isotropic weak coupling BCS value of 1.43. This slightly reduced value could arise from gap anisotropy, or the presence of a small non-superconducting fraction. The superconducting contribution to the entropy $S_{sc}$  can be expressed as \cite{tinkham2004}
\begin{equation}\label{equation8}
S_{sc}(T)=-\frac{3\gamma_n}{k_B\pi^3}\int_{0}^{2\pi}\int_{0}^{\infty}[(1-f)\mathrm{ln}(1-f)+f\mathrm{ln}f]\mathrm{d}\epsilon\mathrm{d}\phi
\end{equation}
where $f$ = [1+exp$(E/k_BT)]^{-1}$ is the Fermi-Dirac function with $E$ = $\sqrt{\epsilon^2+\Delta^2(T)}$, and  $\Delta(T)$ is the superconducting gap, given by $\Delta(T)$ = $\Delta(0)\tanh{1.82[1.018(T_c/T-1)]^{0.51}}$. The $C_{e}/T$ vs $T$ curve in the superconducting state can be well fitted by the single gap $s$-wave model with $\Delta(0)=1.80~k_BT_c$, after taking into account a small non-superconducting fraction. The gap size is slightly larger than the isotropic weak coupling value from the BCS theory. Therefore, the specific heat results indicate that LaPdIn is a conventional superconductor with a single $s$-wave gap.

\subsection{\uppercase\expandafter{c}. In-field resistivity and specific heat}

  \begin{figure}
  	\includegraphics[angle=0,width=0.49\textwidth]{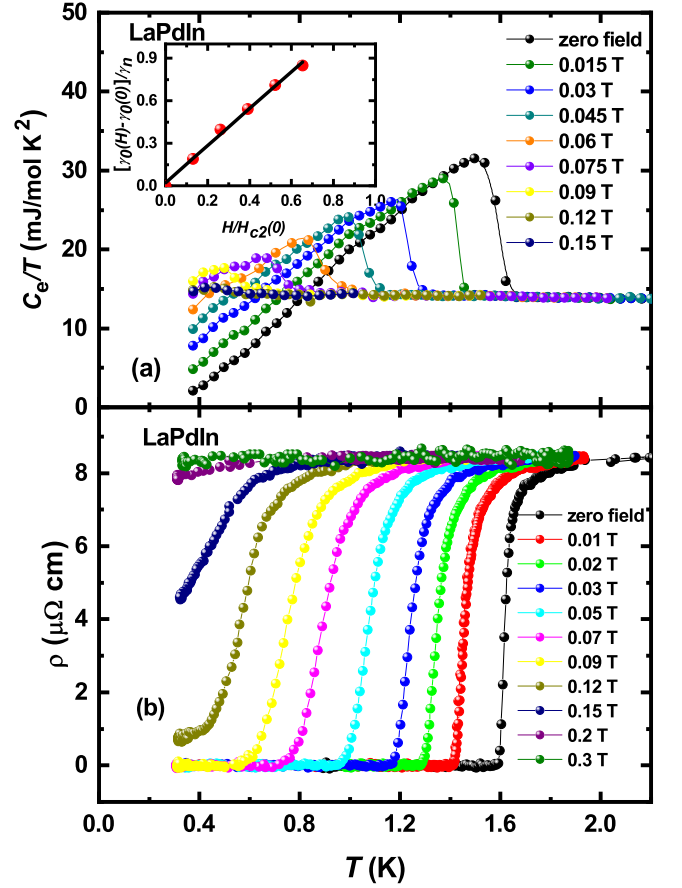}
  	\vspace{-12pt} \caption{\label{Figure4}(Color online)  Temperature dependence of (a) the electronic specific heat as $C_e(T)/T$ and (b) the resistivity $\rho(T)$ of LaPdIn in various applied magnetic fields. The inset of panel (a) shows the field dependence of the residual Sommerfeld coefficient $\gamma_0$, plotted as $[\gamma_0(H)-\gamma_0(0)]/\gamma_n$ versus $H/H_{c2}(0)$. }
  	\vspace{-12pt}
  \end{figure}

  \begin{figure}
  	\includegraphics[angle=0,width=0.49\textwidth]{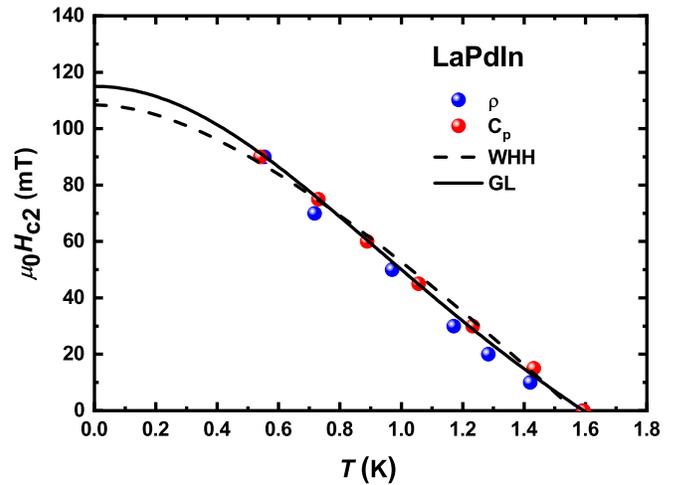}
  	\vspace{-12pt} \caption{\label{Figure5}(Color online) Temperature dependence of the upper critical field $\mu_0H_{c2}$ of LaPdIn. $T_c$ was determined from resistivity and specific heat results. The dashed and solid lines correspond to fitting with the WHH and Ginzburg-Landau models, respectively. }
  	\vspace{-12pt}
  \end{figure}

  \begin{figure}
  	\includegraphics[angle=0,width=0.49\textwidth]{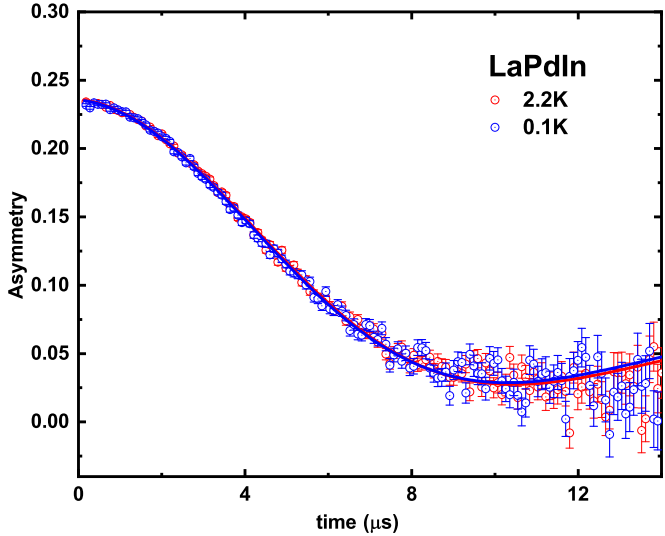}
  	\vspace{-12pt} \caption{\label{Figure6}(Color online)   Zero-field $\mu$SR spectra of LaPdIn at 2.2 K in the normal state (red), and at 0.1 K in the superconducting state (blue). The solid lines show the results of fitting using Eq. \ref{equation10}.  }
  	\vspace{-12pt}
  \end{figure}

    \begin{figure}
  	\includegraphics[angle=0,width=0.49\textwidth]{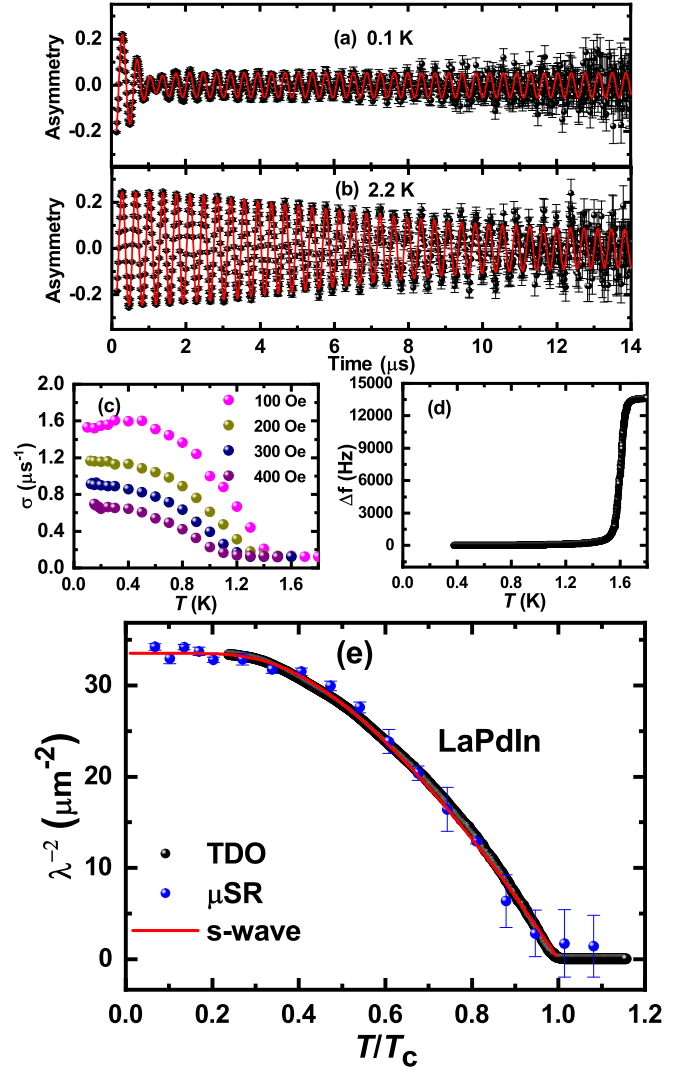}
  	\vspace{-12pt} \caption{\label{Figure7}(Color online) Transverse-field $\mu$SR spectra of LaPdIn in an applied field of 200~Oe at (a) 0.1~K and (b) 2.2~K. The solid lines show the results from fitting with Eq. \ref{equation11}. (c) Temperature dependence of the Gaussian depolarization rate $\sigma$ in different applied transverse fields. (d) TDO frequency shift $\Delta f$ of LaPdIn as a function of temperature from 1.8~K to the lowest measured temperature 0.35~K. (e) Temperature dependence of the London penetration depth  of LaPdIn as $\lambda^{-2}$(T). The blue symbols are derived from the analysis of the TF-$\mu$SR results using Eq.~\ref{equation11}, while the black symbols are from TDO data. The solid line shows the fit to the TF-$\mu$SR data using a single gap $s$-wave model, as described in the text.      }
  	\vspace{-12pt}
  \end{figure}

\begin{table*}[tb]
\caption{Superconducting and normal-state parameters of LaPdIn.}
\label{table:table2}
\begin{ruledtabular}
 \begin{tabular}{c c c c c c c c c c c}
$T_c$   &$H_{c2(GL)}(0)$   &$\lambda(0)$   &$\xi_{GL}(0)$   &$\kappa$    &$\Delta(0)/k_BT_c$    &$\gamma_n$   &$\beta$   &$\lambda_{\rm el-ph}$    &$\Delta C_e/\gamma_nT_c$   &$\Theta_D$  \\
{(K)} &{(mT)} &{(nm)} &{(nm)} &{}  &{} &{(mJ~mol$^{-1}$~K$^{-2}$)} &{(mJ~mol$^{-1}$~K$^{-4}$)} &{} &{} &{(K)}                                \\
\hline\\[-2ex]
{1.59} &{115} &{172.7(8)} &{53.5} &{3.23} &{1.78(6)} &{14.46(6)} &{0.777(5)} &{$0.48-0.57$} &{1.23} &{195(2)} \\
\end{tabular}
\end{ruledtabular}
\end{table*}

  \begin{figure*}
  	\includegraphics[angle=0,width=1\textwidth]{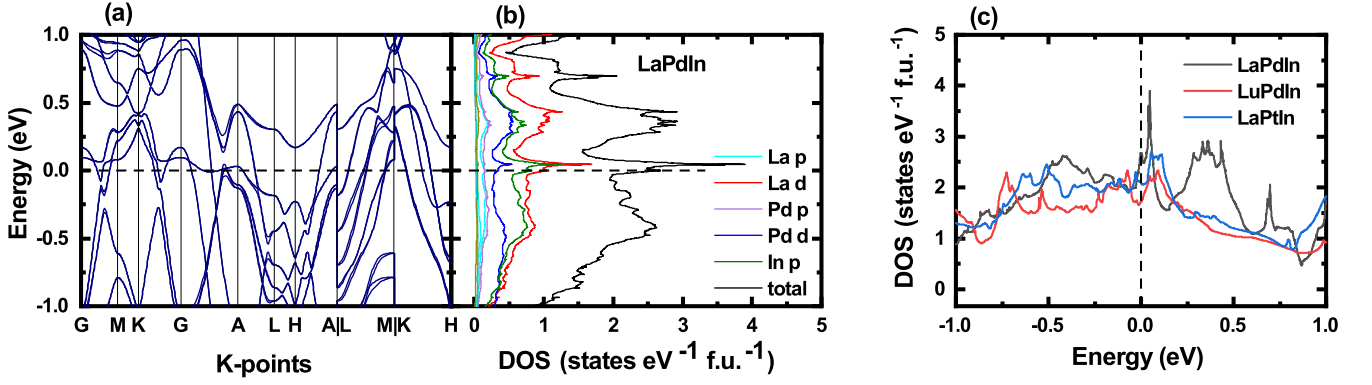}
  	\vspace{-12pt} \caption{\label{Figure8}(Color online)   Results from DFT calculations for LaPdIn showing (a) the band structure within 1~eV of the Fermi level, including spin-orbit coupling. (b) Total and partial density of states of LaPdIn for the calculations including spin-orbit coupling. (c) Density of states of LaPdIn, LuPdIn and LaPtIn within $\pm1$~eV of the Fermi energy level. }
  	\vspace{-12pt}
  \end{figure*}

  \begin{figure}
  	\includegraphics[angle=0,width=0.49\textwidth]{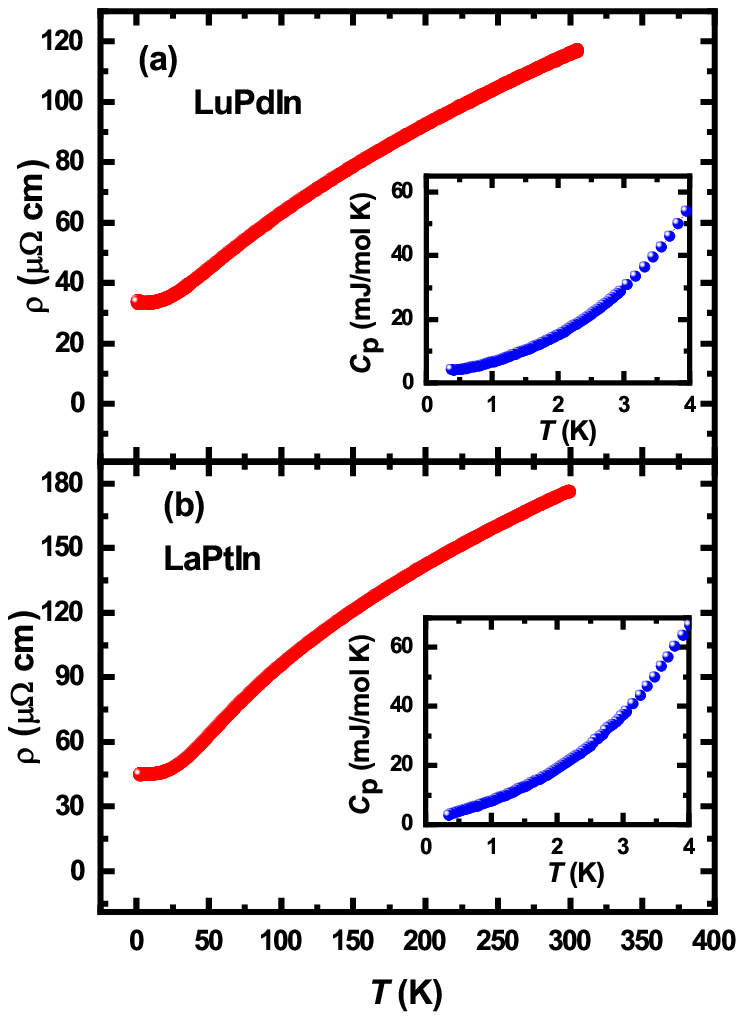}
  	\vspace{-12pt} \caption{\label{Figure9}(Color online)  Temperature dependence of the resistivity and heat capacity of (a) LuPdIn and (b) LaPtIn. The main panels show the resistivity measured from 0.5~K to 300~K. The insets show the low temperature heat capacity. }
  	\vspace{-12pt}
  \end{figure}

Figures \ref{Figure4}(a) and (b) show the temperature dependence of the resistivity and electronic specific heat under various magnetic fields, respectively. In both cases, the transition is suppressed to lower temperature and becomes broader with increasing magnetic field. When a field of 1200~Oe is applied, no transition can be observed in the specific heat, and zero resistivity is not reached at the lowest measured temperature. In the specific heat, the jump at the transition becomes smaller with increasing field, suggesting that it remains second-order in field, characteristic of type-II superconductivity. Figure \ref{Figure5} shows the temperature dependence of the upper critical field $\mu_0H_{c2}$ of LaPdIn. Here the resistivity points are taken from where there is zero resistivity, while for the specific heat, the midpoint of the transition was utilized, and there is good agreement between the two methods. The Werthamer, Helfand, and Hohenberg (WHH) theory \cite{Werthamer1966}, can account for upper critical field using a value of $(d\mu_0H_{c2}/dT)_{T = T_c}$= -88.7 mT/K, yielding a zero temperature upper critical field $\mu_0H_{c2}^{WHH}(0)$ of 108~mT. The Ginzburg-Landau(GL) model was also used,
\begin{equation}\label{equation9}
\mu_0H_{c2}^{GL}(T)=\mu_0H_{c2}^{GL}(0)\frac{1-t^2}{1+t^2}
\end{equation}
where $t$ is the reduced temperature $T/T_c$. $\mu_0H_{c2}^{GL}(0)$ is determined to be 115~mT, far less than the Pauli limit $\mu_0H_{c2}^{Pauli}=1.86~T_c=2.95$~T. The Ginzburg-Landau coherence length $\xi_{GL}$ was calculated using $\mu_0H_{c2}^{GL}(0)=\Phi_0/2\pi\xi_{GL}^2$ to be 53.5~nm, where $\Phi_0$ is the magnetic flux quantum.

From the specific heat measurements under various fields, the residual electronic specific heat coefficient $\gamma_0(H)$ can give information about the low-energy quasiparticle excitations near the Abrikosov vortex cores and thus characterize the superconducting pairing state \cite{Caroli1964}. The inset of Fig. \ref{Figure4}(a) displays the magnetic field dependence of the normalized residual Sommerfeld coefficient $[\gamma_0(H)-\gamma_0(0)]/\gamma_n$, where the linear behavior is consistent with $s$-wave superconductivity. Various parameters corresponding to the properties of the normal and superconducting states of LaPdIn are given  in Table~\ref{table:table2}.

\subsection{\uppercase\expandafter{d}. $\mu$SR and TDO}

$\mu$SR measurements were carried out to investigate the superconducting state of LaPdIn. Zero field (ZF) $\mu$SR spectra are shown at two temperatures in Fig. \ref{Figure6}, in the superconducting state at 0.1~K, and in the normal state at 2.2~K. The data were fitted using the product of the Kubo-Toyabe function with a Lorentzian relaxation,
\begin{equation}\label{equation10}
G_{ZF}(t)=A\left[\frac{1}{3}+\frac{2}{3}(1-\sigma_{ZF}^2t^2)e^{\sigma_{ZF}^2t^2/2}\right]e^{-\Lambda t}
\end{equation}
where $A$ is the initial asymmetry, $\sigma_{ZF}$ and $\Lambda$ are the Kubo-Toyabe and Lorentzian relaxation rates. At 0.1~K, values of $\sigma_{ZF}=0.168(1)~\mu s^{-1}$, $\Lambda=0.014(2)~\mu s^{-1}$ are obtained, while at 2.2~K, $\sigma_{ZF}=0.166(1)~\mu s^{-1}$, $\Lambda=0.015(2)~\mu s^{-1}$. Since the values are very similar above and below $T_c$, this indicates that time-reversal symmetry is preserved in the superconducting state of LaPdIn.

Transverse field (TF) $\mu$SR measurements were performed in several magnetic fields from 100~Oe to 400~Oe. Figures \ref{Figure7}(a) and (b) show spectra obtained at 0.1~K and 2.2~K in a field of 200~Oe, respectively. The asymmetries are described by
\begin{equation}\label{equation11}
A_{TF}(t)=A_se^{-\sigma^2t^2/2}\cos(\gamma_{\mu}B_st+\phi)+A_{bg}\cos(\gamma_{\mu}B_{bg}t+\phi)
\end{equation}
where A$_s$ and A$_{bg}$ are the amplitudes of the asymmetries corresponding to the sample and background, respectively, and $\gamma_{\mu}/2\pi=135.53$~MHz/T is the muon gyromagnetic ratio. B$_s$ and B$_{bg}$ are the local fields corresponding to muons stopping in the sample and background, respectively, while $\phi$ is a common phase and $\sigma$ is the Gaussian depolarization rate. The increase of the depolarization rate upon entering the superconducting state indicates the onset of bulk superconductivity. The temperature dependence of $\sigma$ obtained in various fields is displayed in Fig. \ref{Figure7}(c). The superconducting contribution to the relaxation $\sigma_{sc}$ was obtained after subtracting the nuclear contribution $\sigma_{nuc}$ = 0.121(1)~$\mu s^{-1}$ using $\sigma_{sc}=\sqrt{\sigma^2-\sigma_{nuc}^2}$. $\sigma_{sc}$ at different fields for a given temperature were analyzed using \cite{Shiroka2011}
\begin{equation}\label{equation12}
\sigma_{sc}=4.83\times10^4\frac{\kappa^2(1-b)}{\lambda^2(\kappa^2-0.069)}
\end{equation}
where $\kappa$ is the Ginzburg-Landau parameter, $\lambda$ is the penetration depth, and $b = H/H_{c2}$. The temperature dependence of $\lambda^{-2}(T)$ derived from this analysis is displayed in Fig.~\ref{Figure7}(e), which is proportional to the superfluid density $\rho_s$. Below 0.4~K, $\lambda^{-2}(T)$ is nearly constant, indicating a fully gapped superconducting order parameter, due to the lack of low energy excitations. The  $\lambda^{-2}(T)$  data derived from TF-$\mu$SR are well described by a single gap $s$-wave model
\begin{equation}\label{equation13}
\frac{\lambda^{-2}(T)}{\lambda^{-2}(0)}=1+2\int_{\Delta(T)}^{\infty}\frac{E}{\sqrt{E^2-\Delta^2(T)}}\frac{\partial f}{\partial E}\mathrm{d}E
\end{equation}
where $\Delta(T)$ is the same as used in the specific heat analysis. The solid line shows the results of fitting with such a model, yielding $\Delta(0)$ = 0.245(8)~meV, corresponding to $1.78(6)k_BT_c$, and $\lambda(0)$ = 172.7(8)~nm. With $\xi_{GL}$ = 53.5~nm obtained from the $H_{c2}$ analysis described above, a Ginzburg-Landau parameter ($\kappa$) of 3.23 is obtained using $\kappa=\lambda(0)/\xi_{GL}(0)$, which is larger than $1/\sqrt{2}$, indicating type-II superconductivity in LaPdIn. Note that a slightly smaller value of $\kappa$ = 2.46 is obtained if the value of $\xi_{GL}(0)$ from the $\mu$SR analysis is used, which is likely due to the uncertainties in determining the upper critical field from measurements in relatively low transverse fields.

The TDO method was also utilized to probe the gap structure of LaPdIn. The temperature dependence of the magnetic penetration depth change $\Delta\lambda(T)$ in Fig. \ref{Figure7}(d) shows a sharp superconducting transition, with a $T_c$ consistent with the resistivity and specific heat results. Using the value of $\lambda(0)$ estimated from TF-$\mu$SR, $\lambda^{-2}(T)$ derived the from TDO results is also displayed in Fig. \ref{Figure7}(e). It can be seen that the $\lambda^{-2}$(T) estimated using the two methods are highly consistent.

\subsection{\uppercase\expandafter{e}. DFT calculations}

The calculated band structure and electronic density of states of LaPdIn are displayed in Fig. \ref{Figure8}, where the structural parameters in Table. \ref{table:table1} were utilized. Spin-orbit coupling (SOC) was also taken into consideration. From the momentum-dependent energy curves in Fig. \ref{Figure8}(a), several bands cross the Fermi level. The lifting of the band degeneracy due to antisymmetric spin-orbit coupling can be seen, the average band splitting at the Fermi level $E_{ASOC}$ is only around 6.3~meV. Therefore, the weak influence of ASOC on the electronic structure is consistent with the observation of single gap $s$-wave superconductivity. Figure \ref{Figure8}(b) presents the total and partial electronic densities of state (DOS) of LaPdIn. The DOS at $E_F$ is dominated by La-$d$ and In-$p$ orbitals, with a smaller contribution from Pd-$p$. The total calculated DOS at the Fermi level $N(E_F)$  is 2.65~states/eV-f.u from which  a Sommerfeld coefficient of $\gamma_{cal}=6.25$~mJ~mol$^{-1}$~K$^{-2}$ is calculated, which is smaller than the observed value from the specific heat of $\gamma_n=14.46$~mJ~mol$^{-1}$~K$^{-2}$. Using the values of $\gamma_n$ and $\lambda_{el-ph}$,
\begin{equation}\label{equation14}
\gamma_{cal}=\frac{\gamma_n}{1+\lambda_{el-ph}+\lambda_{corr}}
\end{equation}
a value of $\lambda_{corr}$ = 0.7 is obtained, which corresponds to weak electronic correlations in LaPdIn.

\subsection{\uppercase\expandafter{f}. Physical properties of LuPdIn and LaPtIn}

To explore more possible new isostructural superconductors and to examine the effects on superconductivity of tuning the ASOC, we synthesized various other compounds with the ZrNiAl-type structure, including LuPdIn and LaPtIn. Upon cooling down to 0.4~K, no superconducting transition was found in LuPdIn and LaPtIn from resistivity and specific heat measurements, as shown in Fig. \ref{Figure9}. The phonon contribution of both compounds were estimated from fitting the low temperature specific heat, using $C_p=\gamma_n+\beta T^2+\delta T^4$, yielding $\gamma_n$ of 5.70(1)~mJ~mol$^{-1}$~K$^{-2}$ and 7.34(2)~mJ~mol$^{-1}$~K$^{-2}$ and $\beta$ of 0.461(3)~mJ~mol$^{-1}$~K$^{-4}$ and 0.524(6)~mJ~mol$^{-1}$~K$^{-4}$ for LuPdIn and LaPtIn, respectively. From Eq. \ref{equation4}, $\lambda_{el-ph}$ of both two compounds are estimated to be smaller than 0.41. From comparing the DOS of all three compounds in Fig. \ref{Figure8}(c), it can be seen that $N(E_F)$ of LaPdIn is the largest, consistent with this compound having the largest measured $\gamma_n$ value, which may account for bulk superconductivity only being observed in this material.

\section{\uppercase\expandafter{\romannumeral4}. Discussion and summary}

Since LaPdIn is a NCS, the ASOC can lift the electronic spin degeneracy and possibly result in the mixture of spin-singlet and spin-triplet pairing. The $\mu$SR measurements down to low temperature show that LaPdIn is a fully gapped superconductor well described by a single gap $s$-wave model, which is consistent with results from the TDO-method and specific heat. DFT calculations show that there is a relatively small splitting of the bands due to ASOC, which is in line with the apparent lack of singlet-triplet mixing in this compound. In particular, the size of the band splitting by the ASOC compared to $T_c$, $E_{ASOC}$/$k_BT_c=44$, is small compared to many other noncentrosymmetric superconductors \cite{Smidman2017}, whereas  evidence for singlet-triplet mixing  in weakly correlated systems is often associated with relatively large $E_{ASOC}$/$k_BT_c$, notably in  Li$_2$(Pd$_{1-x}$Pt$_x$)$_3$B \cite{Yuan2006,LeePhysRevB2005, YUAN20081138}. Moreover, despite the predictions of singlet-triplet mixing in NCS, many weakly correlated examples exhibit conventional behavior, well described by one fully open gap, such as Re$_3$W \cite{Biswas2012}, BaPtSi$_3$ \cite{Bauer2009PRB}, La(Pt,Rh)Si$_3$ \cite{Smidman2004, AnandPhysRevB2011} and La(Rh,Ir)P \cite{Qi2014}, and the relationship between ASOC strength and the occurrence of unconventional properties remains to be clarified. In order to understand the role played by ASOC on the superconducting properties, it is of interest to examine isostructural compounds upon the substitution for heavier elements. However, we found a lack of bulk superconductivity in the other studied compounds LuPdIn and LaPtIn, likely due to the lower density of states at the Fermi level. On the other hand, the lower value of $T_c$ for LaPdIn compared to the isostructural pnictide counterparts is likely related to differences in the phonon spectrum and electron-phonon coupling strengths, as evidenced by the larger values of $\theta_D$ and $\lambda_{\rm el-ph}$ in the aforementioned compounds. \cite{das2021probing, BaPhysRevB2019, keiber1984phonon}. Meanwhile, in the centrosymmetric orthorhombic superconductors HfIrSi and ZrIrSi, strongly coupled superconductivity does not appear to be associated with higher values of $T_c$ \cite{Bhattacharyya2019JPCM, PandaPhysRevB2019}. Furthermore, ZF-$\mu$SR measurements reveal that time -reversal symmetry is preserved in the superconducting state of LaPdIn, in line with a number of other NCS.

In summary, we characterized the superconducting order parameter of the noncentrosymmetric superconductor LaPdIn, with $T_c$ = 1.6~K. Specific heat, TF-$\mu$SR, and measurements using the TDO method are all consistent with a single $s$-wave superconducting gap, with a magnitude of 1.8$k_BT_c$ slightly larger than that of weak coupling BCS theory. The upper critical field of 115~mT is much smaller than the Pauli limit, and as such LaPdIn is a type-II superconductor with a relatively low Ginzburg-Landau parameter $\kappa$ = 3.23, and time-reversal symmetry is preserved in the superconducting state. From DFT calculations, the ASOC leads to a relatively small splitting of the bands of around 6~meV, which is considerably less than those NCS exhibiting unconventional properties, which is consistent with the behavior resembling conventional $s$-wave superconductivity.

\section{Acknowledgments}

We thank T. Takabatake for interesting discussions. This work was supported by the National Key R\&D Program of China (No.~2017YFA0303100, No.~2016YFA0300202), the National Natural Science Foundation of China (No. 11874320, No.~12034017 and No.~11974306) and the Key R\&D Program of Zhejiang Province, China (2021C01002).


\begin{thebibliography}{53}%
\makeatletter
\providecommand \@ifxundefined [1]{%
 \@ifx{#1\undefined}
}%
\providecommand \@ifnum [1]{%
 \ifnum #1\expandafter \@firstoftwo
 \else \expandafter \@secondoftwo
 \fi
}%
\providecommand \@ifx [1]{%
 \ifx #1\expandafter \@firstoftwo
 \else \expandafter \@secondoftwo
 \fi
}%
\providecommand \natexlab [1]{#1}%
\providecommand \enquote  [1]{``#1''}%
\providecommand \bibnamefont  [1]{#1}%
\providecommand \bibfnamefont [1]{#1}%
\providecommand \citenamefont [1]{#1}%
\providecommand \href@noop [0]{\@secondoftwo}%
\providecommand \href [0]{\begingroup \@sanitize@url \@href}%
\providecommand \@href[1]{\@@startlink{#1}\@@href}%
\providecommand \@@href[1]{\endgroup#1\@@endlink}%
\providecommand \@sanitize@url [0]{\catcode `\\12\catcode `\$12\catcode
  `\&12\catcode `\#12\catcode `\^12\catcode `\_12\catcode `\%12\relax}%
\providecommand \@@startlink[1]{}%
\providecommand \@@endlink[0]{}%
\providecommand \url  [0]{\begingroup\@sanitize@url \@url }%
\providecommand \@url [1]{\endgroup\@href {#1}{\urlprefix }}%
\providecommand \urlprefix  [0]{URL }%
\providecommand \Eprint [0]{\href }%
\providecommand \doibase [0]{http://dx.doi.org/}%
\providecommand \selectlanguage [0]{\@gobble}%
\providecommand \bibinfo  [0]{\@secondoftwo}%
\providecommand \bibfield  [0]{\@secondoftwo}%
\providecommand \translation [1]{[#1]}%
\providecommand \BibitemOpen [0]{}%
\providecommand \bibitemStop [0]{}%
\providecommand \bibitemNoStop [0]{.\EOS\space}%
\providecommand \EOS [0]{\spacefactor3000\relax}%
\providecommand \BibitemShut  [1]{\csname bibitem#1\endcsname}%
\let\auto@bib@innerbib\@empty
\bibitem [{\citenamefont {Bauer}\ \emph {et~al.}(2004)\citenamefont {Bauer},
  \citenamefont {Hilscher}, \citenamefont {Michor}, \citenamefont {Paul},
  \citenamefont {Scheidt}, \citenamefont {Gribanov}, \citenamefont {Seropegin},
  \citenamefont {No\"el}, \citenamefont {Sigrist},\ and\ \citenamefont
  {Rogl}}]{Bauer2004}%
  \BibitemOpen
  \bibfield  {author} {\bibinfo {author} {\bibfnamefont {E.}~\bibnamefont
  {Bauer}}, \bibinfo {author} {\bibfnamefont {G.}~\bibnamefont {Hilscher}},
  \bibinfo {author} {\bibfnamefont {H.}~\bibnamefont {Michor}}, \bibinfo
  {author} {\bibfnamefont {C.}~\bibnamefont {Paul}}, \bibinfo {author}
  {\bibfnamefont {E.~W.}\ \bibnamefont {Scheidt}}, \bibinfo {author}
  {\bibfnamefont {A.}~\bibnamefont {Gribanov}}, \bibinfo {author}
  {\bibfnamefont {Y.}~\bibnamefont {Seropegin}}, \bibinfo {author}
  {\bibfnamefont {H.}~\bibnamefont {No\"el}}, \bibinfo {author} {\bibfnamefont
  {M.}~\bibnamefont {Sigrist}}, \ and\ \bibinfo {author} {\bibfnamefont
  {P.}~\bibnamefont {Rogl}},\ }\href {\doibase 10.1103/PhysRevLett.92.027003}
  {\bibfield  {journal} {\bibinfo  {journal} {Phys. Rev. Lett.}\ }\textbf
  {\bibinfo {volume} {92}},\ \bibinfo {pages} {027003} (\bibinfo {year}
  {2004})}\BibitemShut {NoStop}%
\bibitem [{\citenamefont {Bauer}\ and\ \citenamefont
  {Sigrist}(2012)}]{bauer2012}%
  \BibitemOpen
  \bibfield  {author} {\bibinfo {author} {\bibfnamefont {E.}~\bibnamefont
  {Bauer}}\ and\ \bibinfo {author} {\bibfnamefont {M.}~\bibnamefont
  {Sigrist}},\ }\href@noop {} {\emph {\bibinfo {title} {Non-centrosymmetric
  superconductors: introduction and overview}}},\ Vol.\ \bibinfo {volume}
  {847}\ (\bibinfo  {publisher} {Springer Science \& Business Media},\ \bibinfo
  {year} {2012})\BibitemShut {NoStop}%
\bibitem [{\citenamefont {Smidman}\ \emph {et~al.}(2017)\citenamefont
  {Smidman}, \citenamefont {Salamon}, \citenamefont {Yuan},\ and\ \citenamefont
  {Agterberg}}]{Smidman2017}%
  \BibitemOpen
  \bibfield  {author} {\bibinfo {author} {\bibfnamefont {M.}~\bibnamefont
  {Smidman}}, \bibinfo {author} {\bibfnamefont {M.~B.}\ \bibnamefont
  {Salamon}}, \bibinfo {author} {\bibfnamefont {H.~Q.}\ \bibnamefont {Yuan}}, \
  and\ \bibinfo {author} {\bibfnamefont {D.~F.}\ \bibnamefont {Agterberg}},\
  }\href {\doibase 10.1088/1361-6633/80/3/036501} {\bibfield  {journal}
  {\bibinfo  {journal} {Reports on Progress in Physics}\ }\textbf {\bibinfo
  {volume} {80}},\ \bibinfo {pages} {036501} (\bibinfo {year}
  {2017})}\BibitemShut {NoStop}%
\bibitem [{\citenamefont {Sato}\ and\ \citenamefont
  {Fujimoto}(2009)}]{SatoPhysRevB2009}%
  \BibitemOpen
  \bibfield  {author} {\bibinfo {author} {\bibfnamefont {M.}~\bibnamefont
  {Sato}}\ and\ \bibinfo {author} {\bibfnamefont {S.}~\bibnamefont
  {Fujimoto}},\ }\href {\doibase 10.1103/PhysRevB.79.094504} {\bibfield
  {journal} {\bibinfo  {journal} {Phys. Rev. B}\ }\textbf {\bibinfo {volume}
  {79}},\ \bibinfo {pages} {094504} (\bibinfo {year} {2009})}\BibitemShut
  {NoStop}%
\bibitem [{\citenamefont {Sato}\ \emph {et~al.}(2010)\citenamefont {Sato},
  \citenamefont {Takahashi},\ and\ \citenamefont
  {Fujimoto}}]{SatoPhysRevB2010}%
  \BibitemOpen
  \bibfield  {author} {\bibinfo {author} {\bibfnamefont {M.}~\bibnamefont
  {Sato}}, \bibinfo {author} {\bibfnamefont {Y.}~\bibnamefont {Takahashi}}, \
  and\ \bibinfo {author} {\bibfnamefont {S.}~\bibnamefont {Fujimoto}},\ }\href
  {\doibase 10.1103/PhysRevB.82.134521} {\bibfield  {journal} {\bibinfo
  {journal} {Phys. Rev. B}\ }\textbf {\bibinfo {volume} {82}},\ \bibinfo
  {pages} {134521} (\bibinfo {year} {2010})}\BibitemShut {NoStop}%
\bibitem [{\citenamefont {Kimura}\ \emph {et~al.}(2005)\citenamefont {Kimura},
  \citenamefont {Ito}, \citenamefont {Saitoh}, \citenamefont {Umeda},
  \citenamefont {Aoki},\ and\ \citenamefont {Terashima}}]{Kimura2005}%
  \BibitemOpen
  \bibfield  {author} {\bibinfo {author} {\bibfnamefont {N.}~\bibnamefont
  {Kimura}}, \bibinfo {author} {\bibfnamefont {K.}~\bibnamefont {Ito}},
  \bibinfo {author} {\bibfnamefont {K.}~\bibnamefont {Saitoh}}, \bibinfo
  {author} {\bibfnamefont {Y.}~\bibnamefont {Umeda}}, \bibinfo {author}
  {\bibfnamefont {H.}~\bibnamefont {Aoki}}, \ and\ \bibinfo {author}
  {\bibfnamefont {T.}~\bibnamefont {Terashima}},\ }\href {\doibase
  10.1103/PhysRevLett.95.247004} {\bibfield  {journal} {\bibinfo  {journal}
  {Phys. Rev. Lett.}\ }\textbf {\bibinfo {volume} {95}},\ \bibinfo {pages}
  {247004} (\bibinfo {year} {2005})}\BibitemShut {NoStop}%
\bibitem [{\citenamefont {Sugitani}\ \emph {et~al.}(2006)\citenamefont
  {Sugitani}, \citenamefont {Okuda}, \citenamefont {Shishido}, \citenamefont
  {Yamada}, \citenamefont {Thamizhavel}, \citenamefont {Yamamoto},
  \citenamefont {D.~Matsuda}, \citenamefont {Haga}, \citenamefont {Takeuchi},
  \citenamefont {Settai},\ and\ \citenamefont {\={O}nuki}}]{Sugitani2006}%
  \BibitemOpen
  \bibfield  {author} {\bibinfo {author} {\bibfnamefont {I.}~\bibnamefont
  {Sugitani}}, \bibinfo {author} {\bibfnamefont {Y.}~\bibnamefont {Okuda}},
  \bibinfo {author} {\bibfnamefont {H.}~\bibnamefont {Shishido}}, \bibinfo
  {author} {\bibfnamefont {T.}~\bibnamefont {Yamada}}, \bibinfo {author}
  {\bibfnamefont {A.}~\bibnamefont {Thamizhavel}}, \bibinfo {author}
  {\bibfnamefont {E.}~\bibnamefont {Yamamoto}}, \bibinfo {author}
  {\bibfnamefont {T.}~\bibnamefont {D.~Matsuda}}, \bibinfo {author}
  {\bibfnamefont {Y.}~\bibnamefont {Haga}}, \bibinfo {author} {\bibfnamefont
  {T.}~\bibnamefont {Takeuchi}}, \bibinfo {author} {\bibfnamefont
  {R.}~\bibnamefont {Settai}}, \ and\ \bibinfo {author} {\bibfnamefont
  {Y.}~\bibnamefont {\={O}nuki}},\ }\href {\doibase 10.1143/JPSJ.75.043703}
  {\bibfield  {journal} {\bibinfo  {journal} {Journal of the Physical Society
  of Japan}\ }\textbf {\bibinfo {volume} {75}},\ \bibinfo {pages} {043703}
  (\bibinfo {year} {2006})}\BibitemShut {NoStop}%
\bibitem [{\citenamefont {Settai}\ \emph {et~al.}(2007)\citenamefont {Settai},
  \citenamefont {Sugitani}, \citenamefont {Okuda}, \citenamefont {Thamizhavel},
  \citenamefont {Nakashima}, \citenamefont {\={O}nuki},\ and\ \citenamefont
  {Harima}}]{Settai2007}%
  \BibitemOpen
  \bibfield  {author} {\bibinfo {author} {\bibfnamefont {R.}~\bibnamefont
  {Settai}}, \bibinfo {author} {\bibfnamefont {I.}~\bibnamefont {Sugitani}},
  \bibinfo {author} {\bibfnamefont {Y.}~\bibnamefont {Okuda}}, \bibinfo
  {author} {\bibfnamefont {A.}~\bibnamefont {Thamizhavel}}, \bibinfo {author}
  {\bibfnamefont {M.}~\bibnamefont {Nakashima}}, \bibinfo {author}
  {\bibfnamefont {Y.}~\bibnamefont {\={O}nuki}}, \ and\ \bibinfo {author}
  {\bibfnamefont {H.}~\bibnamefont {Harima}},\ }\href {\doibase
  https://doi.org/10.1016/j.jmmm.2006.10.717} {\bibfield  {journal} {\bibinfo
  {journal} {Journal of Magnetism and Magnetic Materials}\ }\textbf {\bibinfo
  {volume} {310}},\ \bibinfo {pages} {844 } (\bibinfo {year}
  {2007})}\BibitemShut {NoStop}%
\bibitem [{\citenamefont {Honda}\ \emph {et~al.}(2010)\citenamefont {Honda},
  \citenamefont {Bonalde}, \citenamefont {Yoshiuchi}, \citenamefont {Hirose},
  \citenamefont {Nakamura}, \citenamefont {Shimizu}, \citenamefont {Settai},\
  and\ \citenamefont {\={O}nuki}}]{Honda2010}%
  \BibitemOpen
  \bibfield  {author} {\bibinfo {author} {\bibfnamefont {F.}~\bibnamefont
  {Honda}}, \bibinfo {author} {\bibfnamefont {I.}~\bibnamefont {Bonalde}},
  \bibinfo {author} {\bibfnamefont {S.}~\bibnamefont {Yoshiuchi}}, \bibinfo
  {author} {\bibfnamefont {Y.}~\bibnamefont {Hirose}}, \bibinfo {author}
  {\bibfnamefont {T.}~\bibnamefont {Nakamura}}, \bibinfo {author}
  {\bibfnamefont {K.}~\bibnamefont {Shimizu}}, \bibinfo {author} {\bibfnamefont
  {R.}~\bibnamefont {Settai}}, \ and\ \bibinfo {author} {\bibfnamefont
  {Y.}~\bibnamefont {\={O}nuki}},\ }\href {\doibase
  https://doi.org/10.1016/j.physc.2009.10.126} {\bibfield  {journal} {\bibinfo
  {journal} {Physica C: Superconductivity and its Applications}\ }\textbf
  {\bibinfo {volume} {470}},\ \bibinfo {pages} {S543 } (\bibinfo {year}
  {2010})}\BibitemShut {NoStop}%
\bibitem [{\citenamefont {Akazawa}\ \emph {et~al.}(2004)\citenamefont
  {Akazawa}, \citenamefont {Hidaka}, \citenamefont {Fujiwara}, \citenamefont
  {Kobayashi}, \citenamefont {Yamamoto}, \citenamefont {Haga}, \citenamefont
  {Settai},\ and\ \citenamefont {\={O}nuki}}]{Akazawa2004}%
  \BibitemOpen
  \bibfield  {author} {\bibinfo {author} {\bibfnamefont {T.}~\bibnamefont
  {Akazawa}}, \bibinfo {author} {\bibfnamefont {H.}~\bibnamefont {Hidaka}},
  \bibinfo {author} {\bibfnamefont {T.}~\bibnamefont {Fujiwara}}, \bibinfo
  {author} {\bibfnamefont {T.~C.}\ \bibnamefont {Kobayashi}}, \bibinfo {author}
  {\bibfnamefont {E.}~\bibnamefont {Yamamoto}}, \bibinfo {author}
  {\bibfnamefont {Y.}~\bibnamefont {Haga}}, \bibinfo {author} {\bibfnamefont
  {R.}~\bibnamefont {Settai}}, \ and\ \bibinfo {author} {\bibfnamefont
  {Y.}~\bibnamefont {\={O}nuki}},\ }\href {\doibase 10.1088/0953-8984/16/4/l02}
  {\bibfield  {journal} {\bibinfo  {journal} {Journal of Physics: Condensed
  Matter}\ }\textbf {\bibinfo {volume} {16}},\ \bibinfo {pages} {L29} (\bibinfo
  {year} {2004})}\BibitemShut {NoStop}%
\bibitem [{\citenamefont {Yuan}\ \emph {et~al.}(2006)\citenamefont {Yuan},
  \citenamefont {Agterberg}, \citenamefont {Hayashi}, \citenamefont {Badica},
  \citenamefont {Vandervelde}, \citenamefont {Togano}, \citenamefont
  {Sigrist},\ and\ \citenamefont {Salamon}}]{Yuan2006}%
  \BibitemOpen
  \bibfield  {author} {\bibinfo {author} {\bibfnamefont {H.~Q.}\ \bibnamefont
  {Yuan}}, \bibinfo {author} {\bibfnamefont {D.~F.}\ \bibnamefont {Agterberg}},
  \bibinfo {author} {\bibfnamefont {N.}~\bibnamefont {Hayashi}}, \bibinfo
  {author} {\bibfnamefont {P.}~\bibnamefont {Badica}}, \bibinfo {author}
  {\bibfnamefont {D.}~\bibnamefont {Vandervelde}}, \bibinfo {author}
  {\bibfnamefont {K.}~\bibnamefont {Togano}}, \bibinfo {author} {\bibfnamefont
  {M.}~\bibnamefont {Sigrist}}, \ and\ \bibinfo {author} {\bibfnamefont
  {M.~B.}\ \bibnamefont {Salamon}},\ }\href {\doibase
  10.1103/PhysRevLett.97.017006} {\bibfield  {journal} {\bibinfo  {journal}
  {Phys. Rev. Lett.}\ }\textbf {\bibinfo {volume} {97}},\ \bibinfo {pages}
  {017006} (\bibinfo {year} {2006})}\BibitemShut {NoStop}%
\bibitem [{\citenamefont {Chen}\ \emph {et~al.}(2011)\citenamefont {Chen},
  \citenamefont {Salamon}, \citenamefont {Akutagawa}, \citenamefont {Akimitsu},
  \citenamefont {Singleton}, \citenamefont {Zhang}, \citenamefont {Jiao},\ and\
  \citenamefont {Yuan}}]{Chen2011}%
  \BibitemOpen
  \bibfield  {author} {\bibinfo {author} {\bibfnamefont {J.}~\bibnamefont
  {Chen}}, \bibinfo {author} {\bibfnamefont {M.~B.}\ \bibnamefont {Salamon}},
  \bibinfo {author} {\bibfnamefont {S.}~\bibnamefont {Akutagawa}}, \bibinfo
  {author} {\bibfnamefont {J.}~\bibnamefont {Akimitsu}}, \bibinfo {author}
  {\bibfnamefont {J.}~\bibnamefont {Singleton}}, \bibinfo {author}
  {\bibfnamefont {J.~L.}\ \bibnamefont {Zhang}}, \bibinfo {author}
  {\bibfnamefont {L.}~\bibnamefont {Jiao}}, \ and\ \bibinfo {author}
  {\bibfnamefont {H.~Q.}\ \bibnamefont {Yuan}},\ }\href {\doibase
  10.1103/PhysRevB.83.144529} {\bibfield  {journal} {\bibinfo  {journal} {Phys.
  Rev. B}\ }\textbf {\bibinfo {volume} {83}},\ \bibinfo {pages} {144529}
  (\bibinfo {year} {2011})}\BibitemShut {NoStop}%
\bibitem [{\citenamefont {Adroja}\ \emph {et~al.}(2015)\citenamefont {Adroja},
  \citenamefont {Bhattacharyya}, \citenamefont {Telling}, \citenamefont {Feng},
  \citenamefont {Smidman}, \citenamefont {Pan}, \citenamefont {Zhao},
  \citenamefont {Hillier}, \citenamefont {Pratt},\ and\ \citenamefont
  {Strydom}}]{Adrojaprb2015}%
  \BibitemOpen
  \bibfield  {author} {\bibinfo {author} {\bibfnamefont {D.~T.}\ \bibnamefont
  {Adroja}}, \bibinfo {author} {\bibfnamefont {A.}~\bibnamefont
  {Bhattacharyya}}, \bibinfo {author} {\bibfnamefont {M.}~\bibnamefont
  {Telling}}, \bibinfo {author} {\bibfnamefont {Y.}~\bibnamefont {Feng}},
  \bibinfo {author} {\bibfnamefont {M.}~\bibnamefont {Smidman}}, \bibinfo
  {author} {\bibfnamefont {B.}~\bibnamefont {Pan}}, \bibinfo {author}
  {\bibfnamefont {J.}~\bibnamefont {Zhao}}, \bibinfo {author} {\bibfnamefont
  {A.~D.}\ \bibnamefont {Hillier}}, \bibinfo {author} {\bibfnamefont {F.~L.}\
  \bibnamefont {Pratt}}, \ and\ \bibinfo {author} {\bibfnamefont {A.~M.}\
  \bibnamefont {Strydom}},\ }\href {\doibase 10.1103/PhysRevB.92.134505}
  {\bibfield  {journal} {\bibinfo  {journal} {Phys. Rev. B}\ }\textbf {\bibinfo
  {volume} {92}},\ \bibinfo {pages} {134505} (\bibinfo {year}
  {2015})}\BibitemShut {NoStop}%
\bibitem [{\citenamefont {Pang}\ \emph {et~al.}(2015)\citenamefont {Pang},
  \citenamefont {Smidman}, \citenamefont {Jiang}, \citenamefont {Bao},
  \citenamefont {Weng}, \citenamefont {Wang}, \citenamefont {Jiao},
  \citenamefont {Zhang}, \citenamefont {Cao},\ and\ \citenamefont
  {Yuan}}]{Panggm2015}%
  \BibitemOpen
  \bibfield  {author} {\bibinfo {author} {\bibfnamefont {G.~M.}\ \bibnamefont
  {Pang}}, \bibinfo {author} {\bibfnamefont {M.}~\bibnamefont {Smidman}},
  \bibinfo {author} {\bibfnamefont {W.~B.}\ \bibnamefont {Jiang}}, \bibinfo
  {author} {\bibfnamefont {J.~K.}\ \bibnamefont {Bao}}, \bibinfo {author}
  {\bibfnamefont {Z.~F.}\ \bibnamefont {Weng}}, \bibinfo {author}
  {\bibfnamefont {Y.~F.}\ \bibnamefont {Wang}}, \bibinfo {author}
  {\bibfnamefont {L.}~\bibnamefont {Jiao}}, \bibinfo {author} {\bibfnamefont
  {J.~L.}\ \bibnamefont {Zhang}}, \bibinfo {author} {\bibfnamefont {G.~H.}\
  \bibnamefont {Cao}}, \ and\ \bibinfo {author} {\bibfnamefont {H.~Q.}\
  \bibnamefont {Yuan}},\ }\href {\doibase 10.1103/PhysRevB.91.220502}
  {\bibfield  {journal} {\bibinfo  {journal} {Phys. Rev. B}\ }\textbf {\bibinfo
  {volume} {91}},\ \bibinfo {pages} {220502(R)} (\bibinfo {year}
  {2015})}\BibitemShut {NoStop}%
\bibitem [{\citenamefont {Xie}\ \emph {et~al.}(2020)\citenamefont {Xie},
  \citenamefont {Zhang}, \citenamefont {Shen}, \citenamefont {Jiang},
  \citenamefont {Pang}, \citenamefont {Shang}, \citenamefont {Cao},
  \citenamefont {Smidman},\ and\ \citenamefont {Yuan}}]{Xie2020}%
  \BibitemOpen
  \bibfield  {author} {\bibinfo {author} {\bibfnamefont {W.}~\bibnamefont
  {Xie}}, \bibinfo {author} {\bibfnamefont {P.}~\bibnamefont {Zhang}}, \bibinfo
  {author} {\bibfnamefont {B.}~\bibnamefont {Shen}}, \bibinfo {author}
  {\bibfnamefont {W.}~\bibnamefont {Jiang}}, \bibinfo {author} {\bibfnamefont
  {G.}~\bibnamefont {Pang}}, \bibinfo {author} {\bibfnamefont {T.}~\bibnamefont
  {Shang}}, \bibinfo {author} {\bibfnamefont {C.}~\bibnamefont {Cao}}, \bibinfo
  {author} {\bibfnamefont {M.}~\bibnamefont {Smidman}}, \ and\ \bibinfo
  {author} {\bibfnamefont {H.}~\bibnamefont {Yuan}},\ }\href {\doibase
  10.1007/s11433-019-1488-5} {\bibfield  {journal} {\bibinfo  {journal}
  {Science China Physics, Mechanics and Astronomy}\ }\textbf {\bibinfo {volume}
  {63}},\ \bibinfo {pages} {237412} (\bibinfo {year} {2020})}\BibitemShut
  {NoStop}%
\bibitem [{\citenamefont {Shang}\ \emph {et~al.}(2020)\citenamefont {Shang},
  \citenamefont {Smidman}, \citenamefont {Wang}, \citenamefont {Chang},
  \citenamefont {Baines}, \citenamefont {Lee}, \citenamefont {Nie},
  \citenamefont {Pang}, \citenamefont {Xie}, \citenamefont {Jiang},
  \citenamefont {Shi}, \citenamefont {Medarde}, \citenamefont {Shiroka},\ and\
  \citenamefont {Yuan}}]{Shangprl2020}%
  \BibitemOpen
  \bibfield  {author} {\bibinfo {author} {\bibfnamefont {T.}~\bibnamefont
  {Shang}}, \bibinfo {author} {\bibfnamefont {M.}~\bibnamefont {Smidman}},
  \bibinfo {author} {\bibfnamefont {A.}~\bibnamefont {Wang}}, \bibinfo {author}
  {\bibfnamefont {L.~J.}\ \bibnamefont {Chang}}, \bibinfo {author}
  {\bibfnamefont {C.}~\bibnamefont {Baines}}, \bibinfo {author} {\bibfnamefont
  {M.~K.}\ \bibnamefont {Lee}}, \bibinfo {author} {\bibfnamefont {Z.~Y.}\
  \bibnamefont {Nie}}, \bibinfo {author} {\bibfnamefont {G.~M.}\ \bibnamefont
  {Pang}}, \bibinfo {author} {\bibfnamefont {W.}~\bibnamefont {Xie}}, \bibinfo
  {author} {\bibfnamefont {W.~B.}\ \bibnamefont {Jiang}}, \bibinfo {author}
  {\bibfnamefont {M.}~\bibnamefont {Shi}}, \bibinfo {author} {\bibfnamefont
  {M.}~\bibnamefont {Medarde}}, \bibinfo {author} {\bibfnamefont
  {T.}~\bibnamefont {Shiroka}}, \ and\ \bibinfo {author} {\bibfnamefont
  {H.~Q.}\ \bibnamefont {Yuan}},\ }\href {\doibase
  10.1103/PhysRevLett.124.207001} {\bibfield  {journal} {\bibinfo  {journal}
  {Phys. Rev. Lett.}\ }\textbf {\bibinfo {volume} {124}},\ \bibinfo {pages}
  {207001} (\bibinfo {year} {2020})}\BibitemShut {NoStop}%
\bibitem [{\citenamefont {Bao}\ \emph {et~al.}(2015)\citenamefont {Bao},
  \citenamefont {Liu}, \citenamefont {Ma}, \citenamefont {Meng}, \citenamefont
  {Tang}, \citenamefont {Sun}, \citenamefont {Zhai}, \citenamefont {Jiang},
  \citenamefont {Bai}, \citenamefont {Feng}, \citenamefont {Xu},\ and\
  \citenamefont {Cao}}]{BaoPhysRevX2015}%
  \BibitemOpen
  \bibfield  {author} {\bibinfo {author} {\bibfnamefont {J.-K.}\ \bibnamefont
  {Bao}}, \bibinfo {author} {\bibfnamefont {J.-Y.}\ \bibnamefont {Liu}},
  \bibinfo {author} {\bibfnamefont {C.-W.}\ \bibnamefont {Ma}}, \bibinfo
  {author} {\bibfnamefont {Z.-H.}\ \bibnamefont {Meng}}, \bibinfo {author}
  {\bibfnamefont {Z.-T.}\ \bibnamefont {Tang}}, \bibinfo {author}
  {\bibfnamefont {Y.-L.}\ \bibnamefont {Sun}}, \bibinfo {author} {\bibfnamefont
  {H.-F.}\ \bibnamefont {Zhai}}, \bibinfo {author} {\bibfnamefont
  {H.}~\bibnamefont {Jiang}}, \bibinfo {author} {\bibfnamefont
  {H.}~\bibnamefont {Bai}}, \bibinfo {author} {\bibfnamefont {C.-M.}\
  \bibnamefont {Feng}}, \bibinfo {author} {\bibfnamefont {Z.-A.}\ \bibnamefont
  {Xu}}, \ and\ \bibinfo {author} {\bibfnamefont {G.-H.}\ \bibnamefont {Cao}},\
  }\href {\doibase 10.1103/PhysRevX.5.011013} {\bibfield  {journal} {\bibinfo
  {journal} {Phys. Rev. X}\ }\textbf {\bibinfo {volume} {5}},\ \bibinfo {pages}
  {011013} (\bibinfo {year} {2015})}\BibitemShut {NoStop}%
\bibitem [{\citenamefont {Kong}\ \emph {et~al.}(2015)\citenamefont {Kong},
  \citenamefont {Bud'ko},\ and\ \citenamefont {Canfield}}]{KongPhysRevB2015}%
  \BibitemOpen
  \bibfield  {author} {\bibinfo {author} {\bibfnamefont {T.}~\bibnamefont
  {Kong}}, \bibinfo {author} {\bibfnamefont {S.~L.}\ \bibnamefont {Bud'ko}}, \
  and\ \bibinfo {author} {\bibfnamefont {P.~C.}\ \bibnamefont {Canfield}},\
  }\href {\doibase 10.1103/PhysRevB.91.020507} {\bibfield  {journal} {\bibinfo
  {journal} {Phys. Rev. B}\ }\textbf {\bibinfo {volume} {91}},\ \bibinfo
  {pages} {020507(R)} (\bibinfo {year} {2015})}\BibitemShut {NoStop}%
\bibitem [{\citenamefont {Hillier}\ \emph {et~al.}(2009)\citenamefont
  {Hillier}, \citenamefont {Quintanilla},\ and\ \citenamefont
  {Cywinski}}]{Hillier}%
  \BibitemOpen
  \bibfield  {author} {\bibinfo {author} {\bibfnamefont {A.~D.}\ \bibnamefont
  {Hillier}}, \bibinfo {author} {\bibfnamefont {J.}~\bibnamefont
  {Quintanilla}}, \ and\ \bibinfo {author} {\bibfnamefont {R.}~\bibnamefont
  {Cywinski}},\ }\href {\doibase 10.1103/PhysRevLett.102.117007} {\bibfield
  {journal} {\bibinfo  {journal} {Phys. Rev. Lett.}\ }\textbf {\bibinfo
  {volume} {102}},\ \bibinfo {pages} {117007} (\bibinfo {year}
  {2009})}\BibitemShut {NoStop}%
\bibitem [{\citenamefont {Barker}\ \emph {et~al.}(2015)\citenamefont {Barker},
  \citenamefont {Singh}, \citenamefont {Thamizhavel}, \citenamefont {Hillier},
  \citenamefont {Lees}, \citenamefont {Balakrishnan}, \citenamefont {Paul},\
  and\ \citenamefont {Singh}}]{Barker2015}%
  \BibitemOpen
  \bibfield  {author} {\bibinfo {author} {\bibfnamefont {J.~A.~T.}\
  \bibnamefont {Barker}}, \bibinfo {author} {\bibfnamefont {D.}~\bibnamefont
  {Singh}}, \bibinfo {author} {\bibfnamefont {A.}~\bibnamefont {Thamizhavel}},
  \bibinfo {author} {\bibfnamefont {A.~D.}\ \bibnamefont {Hillier}}, \bibinfo
  {author} {\bibfnamefont {M.~R.}\ \bibnamefont {Lees}}, \bibinfo {author}
  {\bibfnamefont {G.}~\bibnamefont {Balakrishnan}}, \bibinfo {author}
  {\bibfnamefont {D.~M.}\ \bibnamefont {Paul}}, \ and\ \bibinfo {author}
  {\bibfnamefont {R.~P.}\ \bibnamefont {Singh}},\ }\href {\doibase
  10.1103/PhysRevLett.115.267001} {\bibfield  {journal} {\bibinfo  {journal}
  {Phys. Rev. Lett.}\ }\textbf {\bibinfo {volume} {115}},\ \bibinfo {pages}
  {267001} (\bibinfo {year} {2015})}\BibitemShut {NoStop}%
\bibitem [{\citenamefont {Singh}\ \emph {et~al.}(2020)\citenamefont {Singh},
  \citenamefont {Scheurer}, \citenamefont {Hillier}, \citenamefont {Adroja},\
  and\ \citenamefont {Singh}}]{SinghPhysRevB2020}%
  \BibitemOpen
  \bibfield  {author} {\bibinfo {author} {\bibfnamefont {D.}~\bibnamefont
  {Singh}}, \bibinfo {author} {\bibfnamefont {M.~S.}\ \bibnamefont {Scheurer}},
  \bibinfo {author} {\bibfnamefont {A.~D.}\ \bibnamefont {Hillier}}, \bibinfo
  {author} {\bibfnamefont {D.~T.}\ \bibnamefont {Adroja}}, \ and\ \bibinfo
  {author} {\bibfnamefont {R.~P.}\ \bibnamefont {Singh}},\ }\href {\doibase
  10.1103/PhysRevB.102.134511} {\bibfield  {journal} {\bibinfo  {journal}
  {Phys. Rev. B}\ }\textbf {\bibinfo {volume} {102}},\ \bibinfo {pages}
  {134511} (\bibinfo {year} {2020})}\BibitemShut {NoStop}%
\bibitem [{\citenamefont {Singh}\ \emph {et~al.}(2014)\citenamefont {Singh},
  \citenamefont {Hillier}, \citenamefont {Mazidian}, \citenamefont
  {Quintanilla}, \citenamefont {Annett}, \citenamefont {Paul}, \citenamefont
  {Balakrishnan},\ and\ \citenamefont {Lees}}]{Singh2014}%
  \BibitemOpen
  \bibfield  {author} {\bibinfo {author} {\bibfnamefont {R.~P.}\ \bibnamefont
  {Singh}}, \bibinfo {author} {\bibfnamefont {A.~D.}\ \bibnamefont {Hillier}},
  \bibinfo {author} {\bibfnamefont {B.}~\bibnamefont {Mazidian}}, \bibinfo
  {author} {\bibfnamefont {J.}~\bibnamefont {Quintanilla}}, \bibinfo {author}
  {\bibfnamefont {J.~F.}\ \bibnamefont {Annett}}, \bibinfo {author}
  {\bibfnamefont {D.~M.}\ \bibnamefont {Paul}}, \bibinfo {author}
  {\bibfnamefont {G.}~\bibnamefont {Balakrishnan}}, \ and\ \bibinfo {author}
  {\bibfnamefont {M.~R.}\ \bibnamefont {Lees}},\ }\href {\doibase
  10.1103/PhysRevLett.112.107002} {\bibfield  {journal} {\bibinfo  {journal}
  {Phys. Rev. Lett.}\ }\textbf {\bibinfo {volume} {112}},\ \bibinfo {pages}
  {107002} (\bibinfo {year} {2014})}\BibitemShut {NoStop}%
\bibitem [{\citenamefont {Shang}\ \emph {et~al.}(2018)\citenamefont {Shang},
  \citenamefont {Smidman}, \citenamefont {Ghosh}, \citenamefont {Baines},
  \citenamefont {Chang}, \citenamefont {Gawryluk}, \citenamefont {Barker},
  \citenamefont {Singh}, \citenamefont {Paul}, \citenamefont {Balakrishnan},
  \citenamefont {Pomjakushina}, \citenamefont {Shi}, \citenamefont {Medarde},
  \citenamefont {Hillier}, \citenamefont {Yuan}, \citenamefont {Quintanilla},
  \citenamefont {Mesot},\ and\ \citenamefont {Shiroka}}]{Shangprl2018}%
  \BibitemOpen
  \bibfield  {author} {\bibinfo {author} {\bibfnamefont {T.}~\bibnamefont
  {Shang}}, \bibinfo {author} {\bibfnamefont {M.}~\bibnamefont {Smidman}},
  \bibinfo {author} {\bibfnamefont {S.~K.}\ \bibnamefont {Ghosh}}, \bibinfo
  {author} {\bibfnamefont {C.}~\bibnamefont {Baines}}, \bibinfo {author}
  {\bibfnamefont {L.~J.}\ \bibnamefont {Chang}}, \bibinfo {author}
  {\bibfnamefont {D.~J.}\ \bibnamefont {Gawryluk}}, \bibinfo {author}
  {\bibfnamefont {J.~A.~T.}\ \bibnamefont {Barker}}, \bibinfo {author}
  {\bibfnamefont {R.~P.}\ \bibnamefont {Singh}}, \bibinfo {author}
  {\bibfnamefont {D.~M.}\ \bibnamefont {Paul}}, \bibinfo {author}
  {\bibfnamefont {G.}~\bibnamefont {Balakrishnan}}, \bibinfo {author}
  {\bibfnamefont {E.}~\bibnamefont {Pomjakushina}}, \bibinfo {author}
  {\bibfnamefont {M.}~\bibnamefont {Shi}}, \bibinfo {author} {\bibfnamefont
  {M.}~\bibnamefont {Medarde}}, \bibinfo {author} {\bibfnamefont {A.~D.}\
  \bibnamefont {Hillier}}, \bibinfo {author} {\bibfnamefont {H.~Q.}\
  \bibnamefont {Yuan}}, \bibinfo {author} {\bibfnamefont {J.}~\bibnamefont
  {Quintanilla}}, \bibinfo {author} {\bibfnamefont {J.}~\bibnamefont {Mesot}},
  \ and\ \bibinfo {author} {\bibfnamefont {T.}~\bibnamefont {Shiroka}},\ }\href
  {\doibase 10.1103/PhysRevLett.121.257002} {\bibfield  {journal} {\bibinfo
  {journal} {Phys. Rev. Lett.}\ }\textbf {\bibinfo {volume} {121}},\ \bibinfo
  {pages} {257002} (\bibinfo {year} {2018})}\BibitemShut {NoStop}%
\bibitem [{\citenamefont {Singh}\ \emph {et~al.}(2017)\citenamefont {Singh},
  \citenamefont {Barker}, \citenamefont {Thamizhavel}, \citenamefont {Paul},
  \citenamefont {Hillier},\ and\ \citenamefont {Singh}}]{SinghPhysRevB2017}%
  \BibitemOpen
  \bibfield  {author} {\bibinfo {author} {\bibfnamefont {D.}~\bibnamefont
  {Singh}}, \bibinfo {author} {\bibfnamefont {J.~A.~T.}\ \bibnamefont
  {Barker}}, \bibinfo {author} {\bibfnamefont {A.}~\bibnamefont {Thamizhavel}},
  \bibinfo {author} {\bibfnamefont {D.~M.}\ \bibnamefont {Paul}}, \bibinfo
  {author} {\bibfnamefont {A.~D.}\ \bibnamefont {Hillier}}, \ and\ \bibinfo
  {author} {\bibfnamefont {R.~P.}\ \bibnamefont {Singh}},\ }\href {\doibase
  10.1103/PhysRevB.96.180501} {\bibfield  {journal} {\bibinfo  {journal} {Phys.
  Rev. B}\ }\textbf {\bibinfo {volume} {96}},\ \bibinfo {pages} {180501(R)}
  (\bibinfo {year} {2017})}\BibitemShut {NoStop}%
\bibitem [{\citenamefont {Singh}\ \emph {et~al.}(2018)\citenamefont {Singh},
  \citenamefont {K.~P.}, \citenamefont {Barker}, \citenamefont {Paul},
  \citenamefont {Hillier},\ and\ \citenamefont {Singh}}]{SinghPhysRevB2018}%
  \BibitemOpen
  \bibfield  {author} {\bibinfo {author} {\bibfnamefont {D.}~\bibnamefont
  {Singh}}, \bibinfo {author} {\bibfnamefont {Sajilesh}~\bibnamefont {K.~P.}},
  \bibinfo {author} {\bibfnamefont {J.~A.~T.}\ \bibnamefont {Barker}}, \bibinfo
  {author} {\bibfnamefont {D.~M.}\ \bibnamefont {Paul}}, \bibinfo {author}
  {\bibfnamefont {A.~D.}\ \bibnamefont {Hillier}}, \ and\ \bibinfo {author}
  {\bibfnamefont {R.~P.}\ \bibnamefont {Singh}},\ }\href {\doibase
  10.1103/PhysRevB.97.100505} {\bibfield  {journal} {\bibinfo  {journal} {Phys.
  Rev. B}\ }\textbf {\bibinfo {volume} {97}},\ \bibinfo {pages} {100505(R)}
  (\bibinfo {year} {2018})}\BibitemShut {NoStop}%
\bibitem [{\citenamefont {Zumdick}\ \emph {et~al.}(1999)\citenamefont
  {Zumdick}, \citenamefont {Hoffmann},\ and\ \citenamefont
  {P?ttgen}}]{Markus1999}%
  \BibitemOpen
  \bibfield  {author} {\bibinfo {author} {\bibfnamefont {M.~F.}\ \bibnamefont
  {Zumdick}}, \bibinfo {author} {\bibfnamefont {R.-D.}\ \bibnamefont
  {Hoffmann}}, \ and\ \bibinfo {author} {\bibfnamefont {R.}~\bibnamefont
  {P{\"o}ttgen}},\ }\href {\doibase https://doi.org/10.1515/znb-1999-0111}
  {\bibfield  {journal} {\bibinfo  {journal} {Zeitschrift f\"{u}r Naturforschung
  B}\ }\textbf {\bibinfo {volume} {54}},\ \bibinfo {pages} {45 } (\bibinfo
  {year} {1999})}\BibitemShut {NoStop}%
\bibitem [{\citenamefont {Barz}\ \emph {et~al.}(1980)\citenamefont {Barz},
  \citenamefont {Ku}, \citenamefont {Meisner}, \citenamefont {Fisk},\ and\
  \citenamefont {Matthias}}]{Barz1980}%
  \BibitemOpen
  \bibfield  {author} {\bibinfo {author} {\bibfnamefont {H.}~\bibnamefont
  {Barz}}, \bibinfo {author} {\bibfnamefont {H.~C.}\ \bibnamefont {Ku}},
  \bibinfo {author} {\bibfnamefont {G.~P.}\ \bibnamefont {Meisner}}, \bibinfo
  {author} {\bibfnamefont {Z.}~\bibnamefont {Fisk}}, \ and\ \bibinfo {author}
  {\bibfnamefont {B.~T.}\ \bibnamefont {Matthias}},\ }\href {\doibase
  10.1073/pnas.77.6.3132} {\bibfield  {journal} {\bibinfo  {journal}
  {Proceedings of the National Academy of Sciences}\ }\textbf {\bibinfo
  {volume} {77}},\ \bibinfo {pages} {3132} (\bibinfo {year}
  {1980})}\BibitemShut {NoStop}%
\bibitem [{\citenamefont {Meisner}\ \emph {et~al.}(1983)\citenamefont
  {Meisner}, \citenamefont {Ku},\ and\ \citenamefont {Barz}}]{MEISNER1983}%
  \BibitemOpen
  \bibfield  {author} {\bibinfo {author} {\bibfnamefont {G.}~\bibnamefont
  {Meisner}}, \bibinfo {author} {\bibfnamefont {H.}~\bibnamefont {Ku}}, \ and\
  \bibinfo {author} {\bibfnamefont {H.}~\bibnamefont {Barz}},\ }\href {\doibase
  https://doi.org/10.1016/0025-5408(83)90010-7} {\bibfield  {journal} {\bibinfo
   {journal} {Materials Research Bulletin}\ }\textbf {\bibinfo {volume} {18}},\
  \bibinfo {pages} {983 } (\bibinfo {year} {1983})}\BibitemShut {NoStop}%
\bibitem [{\citenamefont {Shirotani}\ \emph {et~al.}(1999)\citenamefont
  {Shirotani}, \citenamefont {Tachi}, \citenamefont {Konno}, \citenamefont
  {Todo},\ and\ \citenamefont {Yagi}}]{Ichimin1999}%
  \BibitemOpen
  \bibfield  {author} {\bibinfo {author} {\bibfnamefont {I.}~\bibnamefont
  {Shirotani}}, \bibinfo {author} {\bibfnamefont {K.}~\bibnamefont {Tachi}},
  \bibinfo {author} {\bibfnamefont {Y.}~\bibnamefont {Konno}}, \bibinfo
  {author} {\bibfnamefont {S.}~\bibnamefont {Todo}}, \ and\ \bibinfo {author}
  {\bibfnamefont {T.}~\bibnamefont {Yagi}},\ }\href {\doibase
  10.1080/13642819908205748} {\bibfield  {journal} {\bibinfo  {journal}
  {Philosophical Magazine B}\ }\textbf {\bibinfo {volume} {79}},\ \bibinfo
  {pages} {767} (\bibinfo {year} {1999})}\BibitemShut {NoStop}%
\bibitem [{\citenamefont {Das}\ \emph {et~al.}(2021)\citenamefont {Das},
  \citenamefont {Adroja}, \citenamefont {Lees}, \citenamefont {Taylor},
  \citenamefont {Bishnoi}, \citenamefont {Anand}, \citenamefont
  {Bhattacharyya}, \citenamefont {Guguchia}, \citenamefont {Baines},
  \citenamefont {Luetkens} \emph {et~al.}}]{das2021probing}%
  \BibitemOpen
  \bibfield  {author} {\bibinfo {author} {\bibfnamefont {D.}~\bibnamefont
  {Das}}, \bibinfo {author} {\bibfnamefont {D.~T.}\ \bibnamefont {Adroja}},
  \bibinfo {author} {\bibfnamefont {M.~R.}\ \bibnamefont {Lees}}, \bibinfo
  {author} {\bibfnamefont {R.~W.}\ \bibnamefont {Taylor}}, \bibinfo {author}
  {\bibfnamefont {Z.~S.}\ \bibnamefont {Bishnoi}}, \bibinfo {author}
  {\bibfnamefont {V.~K.}\ \bibnamefont {Anand}}, \bibinfo {author}
  {\bibfnamefont {A.}~\bibnamefont {Bhattacharyya}}, \bibinfo {author}
  {\bibfnamefont {Z.}~\bibnamefont {Guguchia}}, \bibinfo {author}
  {\bibfnamefont {C.}~\bibnamefont {Baines}}, \bibinfo {author} {\bibfnamefont
  {H.}~\bibnamefont {Luetkens}}, \bibinfo {author} {\bibfnamefont {G.~B.~G.}\
  \bibnamefont {Stenning}}, \bibinfo {author} {\bibfnamefont {L.}~\bibnamefont
  {Duan}}, \bibinfo {author} {\bibfnamefont {X.}~\bibnamefont {Wang}}, \ and\
  \bibinfo {author} {\bibfnamefont {C.}~\bibnamefont {Jin}},\ }\href {\doibase
  10.1103/PhysRevB.103.144516} {\bibfield  {journal} {\bibinfo  {journal}
  {Phys. Rev. B}\ }\textbf {\bibinfo {volume} {103}},\ \bibinfo {pages}
  {144516} (\bibinfo {year} {2021})}\BibitemShut {NoStop}%
\bibitem [{\citenamefont {Shirotani}\ \emph {et~al.}(2000)\citenamefont
  {Shirotani}, \citenamefont {Takaya}, \citenamefont {Kaneko}, \citenamefont
  {Sekine},\ and\ \citenamefont {Yagi}}]{SHIROTANI2000}%
  \BibitemOpen
  \bibfield  {author} {\bibinfo {author} {\bibfnamefont {I.}~\bibnamefont
  {Shirotani}}, \bibinfo {author} {\bibfnamefont {M.}~\bibnamefont {Takaya}},
  \bibinfo {author} {\bibfnamefont {I.}~\bibnamefont {Kaneko}}, \bibinfo
  {author} {\bibfnamefont {C.}~\bibnamefont {Sekine}}, \ and\ \bibinfo {author}
  {\bibfnamefont {T.}~\bibnamefont {Yagi}},\ }\href {\doibase
  https://doi.org/10.1016/S0038-1098(00)00393-8} {\bibfield  {journal}
  {\bibinfo  {journal} {Solid State Communications}\ }\textbf {\bibinfo
  {volume} {116}},\ \bibinfo {pages} {683 } (\bibinfo {year}
  {2000})}\BibitemShut {NoStop}%
\bibitem [{\citenamefont {Okamoto}\ \emph {et~al.}(2016)\citenamefont
  {Okamoto}, \citenamefont {Inohara}, \citenamefont {Yamakawa}, \citenamefont
  {Yamakage},\ and\ \citenamefont {Takenaka}}]{Okamoto2016}%
  \BibitemOpen
  \bibfield  {author} {\bibinfo {author} {\bibfnamefont {Y.}~\bibnamefont
  {Okamoto}}, \bibinfo {author} {\bibfnamefont {T.}~\bibnamefont {Inohara}},
  \bibinfo {author} {\bibfnamefont {Y.}~\bibnamefont {Yamakawa}}, \bibinfo
  {author} {\bibfnamefont {A.}~\bibnamefont {Yamakage}}, \ and\ \bibinfo
  {author} {\bibfnamefont {K.}~\bibnamefont {Takenaka}},\ }\href {\doibase
  10.7566/JPSJ.85.013704} {\bibfield  {journal} {\bibinfo  {journal} {Journal
  of the Physical Society of Japan}\ }\textbf {\bibinfo {volume} {85}},\
  \bibinfo {pages} {013704} (\bibinfo {year} {2016})}\BibitemShut {NoStop}%
\bibitem [{\citenamefont {Inohara}\ \emph {et~al.}(2016)\citenamefont
  {Inohara}, \citenamefont {Okamoto}, \citenamefont {Yamakawa},\ and\
  \citenamefont {Takenaka}}]{Inohara2016}%
  \BibitemOpen
  \bibfield  {author} {\bibinfo {author} {\bibfnamefont {T.}~\bibnamefont
  {Inohara}}, \bibinfo {author} {\bibfnamefont {Y.}~\bibnamefont {Okamoto}},
  \bibinfo {author} {\bibfnamefont {Y.}~\bibnamefont {Yamakawa}}, \ and\
  \bibinfo {author} {\bibfnamefont {K.}~\bibnamefont {Takenaka}},\ }\href
  {\doibase 10.7566/JPSJ.85.094706} {\bibfield  {journal} {\bibinfo  {journal}
  {Journal of the Physical Society of Japan}\ }\textbf {\bibinfo {volume}
  {85}},\ \bibinfo {pages} {094706} (\bibinfo {year} {2016})}\BibitemShut
  {NoStop}%
\bibitem [{\citenamefont {Mihalik}\ \emph {et~al.}(2008)\citenamefont
  {Mihalik}, \citenamefont {Sechovsk\'{y}}, \citenamefont {Divi\u{s}}, \citenamefont
  {Gab\'{a}ni},\ and\ \citenamefont {Mihalik}}]{MIHALIK2008}%
  \BibitemOpen
  \bibfield  {author} {\bibinfo {author} {\bibfnamefont {M.}~\bibnamefont
  {Mihalik}}, \bibinfo {author} {\bibfnamefont {V.}~\bibnamefont {Sechovsk{\`y}}},
  \bibinfo {author} {\bibfnamefont {M.}~\bibnamefont {Divi{\v{s}}}}, \bibinfo
  {author} {\bibfnamefont {S.}~\bibnamefont {Gab\'{a}ni}}, \ and\ \bibinfo
  {author} {\bibfnamefont {M.}~\bibnamefont {Mihalik}},\ }\href {\doibase
  https://doi.org/10.1016/j.jallcom.2007.01.121} {\bibfield  {journal}
  {\bibinfo  {journal} {Journal of Alloys and Compounds}\ }\textbf {\bibinfo
  {volume} {452}},\ \bibinfo {pages} {241 } (\bibinfo {year}
  {2008})}\BibitemShut {NoStop}%
\bibitem [{\citenamefont {Br��ck}\ \emph {et~al.}(1988)\citenamefont {Br��ck},
  \citenamefont {van Sprang}, \citenamefont {Klaasse},\ and\ \citenamefont
  {de~Boer}}]{Bruck1988}%
  \BibitemOpen
  \bibfield  {author} {\bibinfo {author} {\bibfnamefont {E.}~\bibnamefont
  {Br\"{u}ck}}, \bibinfo {author} {\bibfnamefont {M.}~\bibnamefont {van Sprang}},
  \bibinfo {author} {\bibfnamefont {J.~C.~P.}\ \bibnamefont {Klaasse}}, \ and\
  \bibinfo {author} {\bibfnamefont {F.~R.}\ \bibnamefont {de~Boer}},\ }\href
  {\doibase 10.1063/1.340751} {\bibfield  {journal} {\bibinfo  {journal}
  {Journal of Applied Physics}\ }\textbf {\bibinfo {volume} {63}},\ \bibinfo
  {pages} {3417} (\bibinfo {year} {1988})}\BibitemShut {NoStop}%
\bibitem [{\citenamefont {Van~Degrift}(1975)}]{Degrift1975}%
  \BibitemOpen
  \bibfield  {author} {\bibinfo {author} {\bibfnamefont {C.~T.}\ \bibnamefont
  {Van~Degrift}},\ }\href {\doibase 10.1063/1.1134272} {\bibfield  {journal}
  {\bibinfo  {journal} {Review of Scientific Instruments}\ }\textbf {\bibinfo
  {volume} {46}},\ \bibinfo {pages} {599} (\bibinfo {year} {1975})}\BibitemShut
  {NoStop}%
\bibitem [{\citenamefont {Prozorov}\ \emph {et~al.}(2000)\citenamefont
  {Prozorov}, \citenamefont {Giannetta}, \citenamefont {Carrington},\ and\
  \citenamefont {Araujo-Moreira}}]{Prozorov2000}%
  \BibitemOpen
  \bibfield  {author} {\bibinfo {author} {\bibfnamefont {R.}~\bibnamefont
  {Prozorov}}, \bibinfo {author} {\bibfnamefont {R.~W.}\ \bibnamefont
  {Giannetta}}, \bibinfo {author} {\bibfnamefont {A.}~\bibnamefont
  {Carrington}}, \ and\ \bibinfo {author} {\bibfnamefont {F.~M.}\ \bibnamefont
  {Araujo-Moreira}},\ }\href {\doibase 10.1103/PhysRevB.62.115} {\bibfield
  {journal} {\bibinfo  {journal} {Phys. Rev. B}\ }\textbf {\bibinfo {volume}
  {62}},\ \bibinfo {pages} {115} (\bibinfo {year} {2000})}\BibitemShut
  {NoStop}%
\bibitem [{\citenamefont {Gondek}\ \emph {et~al.}(2007)\citenamefont {Gondek},
  \citenamefont {Szytu{\l}a}, \citenamefont {Kaczorowski},\ and\ \citenamefont
  {Nenkov}}]{GONDEK2007}%
  \BibitemOpen
  \bibfield  {author} {\bibinfo {author} {\bibfnamefont {{\L}.}~\bibnamefont
  {Gondek}}, \bibinfo {author} {\bibfnamefont {A.}~\bibnamefont {Szytu{\l}a}},
  \bibinfo {author} {\bibfnamefont {D.}~\bibnamefont {Kaczorowski}}, \ and\
  \bibinfo {author} {\bibfnamefont {K.}~\bibnamefont {Nenkov}},\ }\href
  {\doibase https://doi.org/10.1016/j.ssc.2007.04.015} {\bibfield  {journal}
  {\bibinfo  {journal} {Solid State Communications}\ }\textbf {\bibinfo
  {volume} {142}},\ \bibinfo {pages} {556 } (\bibinfo {year}
  {2007})}\BibitemShut {NoStop}%
\bibitem [{\citenamefont {Tinkham}(2004)}]{tinkham2004}%
  \BibitemOpen
  \bibfield  {author} {\bibinfo {author} {\bibfnamefont {M.}~\bibnamefont
  {Tinkham}},\ }\href@noop {} {\emph {\bibinfo {title} {Introduction to
  superconductivity}}}\ (\bibinfo  {publisher} {Courier Corporation},\ \bibinfo
  {year} {2004})\BibitemShut {NoStop}%
\bibitem [{\citenamefont {Werthamer}\ \emph {et~al.}(1966)\citenamefont
  {Werthamer}, \citenamefont {Helfand},\ and\ \citenamefont
  {Hohenberg}}]{Werthamer1966}%
  \BibitemOpen
  \bibfield  {author} {\bibinfo {author} {\bibfnamefont {N.~R.}\ \bibnamefont
  {Werthamer}}, \bibinfo {author} {\bibfnamefont {E.}~\bibnamefont {Helfand}},
  \ and\ \bibinfo {author} {\bibfnamefont {P.~C.}\ \bibnamefont {Hohenberg}},\
  }\href {\doibase 10.1103/PhysRev.147.295} {\bibfield  {journal} {\bibinfo
  {journal} {Phys. Rev.}\ }\textbf {\bibinfo {volume} {147}},\ \bibinfo {pages}
  {295} (\bibinfo {year} {1966})}\BibitemShut {NoStop}%
\bibitem [{\citenamefont {Caroli}\ \emph {et~al.}(1964)\citenamefont {Caroli},
  \citenamefont {{De Gennes}},\ and\ \citenamefont {Matricon}}]{Caroli1964}%
  \BibitemOpen
  \bibfield  {author} {\bibinfo {author} {\bibfnamefont {C.}~\bibnamefont
  {Caroli}}, \bibinfo {author} {\bibfnamefont {P.}~\bibnamefont {{De Gennes}}},
  \ and\ \bibinfo {author} {\bibfnamefont {J.}~\bibnamefont {Matricon}},\
  }\href {\doibase https://doi.org/10.1016/0031-9163(64)90375-0} {\bibfield
  {journal} {\bibinfo  {journal} {Physics Letters}\ }\textbf {\bibinfo {volume}
  {9}},\ \bibinfo {pages} {307 } (\bibinfo {year} {1964})}\BibitemShut
  {NoStop}%
\bibitem [{\citenamefont {Shiroka}\ \emph {et~al.}(2011)\citenamefont
  {Shiroka}, \citenamefont {Lamura}, \citenamefont {Renzi}, \citenamefont
  {Belli}, \citenamefont {Emery}, \citenamefont {Rida}, \citenamefont {Cahen},
  \citenamefont {Mar{\^{e}}ch{\'{e}}}, \citenamefont {Lagrange},\ and\
  \citenamefont {H{\'{e}}rold}}]{Shiroka2011}%
  \BibitemOpen
  \bibfield  {author} {\bibinfo {author} {\bibfnamefont {T.}~\bibnamefont
  {Shiroka}}, \bibinfo {author} {\bibfnamefont {G.}~\bibnamefont {Lamura}},
  \bibinfo {author} {\bibfnamefont {R.~D.}\ \bibnamefont {Renzi}}, \bibinfo
  {author} {\bibfnamefont {M.}~\bibnamefont {Belli}}, \bibinfo {author}
  {\bibfnamefont {N.}~\bibnamefont {Emery}}, \bibinfo {author} {\bibfnamefont
  {H.}~\bibnamefont {Rida}}, \bibinfo {author} {\bibfnamefont {S.}~\bibnamefont
  {Cahen}}, \bibinfo {author} {\bibfnamefont {J.-F.}\ \bibnamefont
  {Mar{\^{e}}ch{\'{e}}}}, \bibinfo {author} {\bibfnamefont {P.}~\bibnamefont
  {Lagrange}}, \ and\ \bibinfo {author} {\bibfnamefont {C.}~\bibnamefont
  {H{\'{e}}rold}},\ }\href {\doibase 10.1088/1367-2630/13/1/013038} {\bibfield
  {journal} {\bibinfo  {journal} {New Journal of Physics}\ }\textbf {\bibinfo
  {volume} {13}},\ \bibinfo {pages} {013038} (\bibinfo {year}
  {2011})}\BibitemShut {NoStop}%
\bibitem [{\citenamefont {Lee}\ and\ \citenamefont
  {Pickett}(2005)}]{LeePhysRevB2005}%
  \BibitemOpen
  \bibfield  {author} {\bibinfo {author} {\bibfnamefont {K.-W.}\ \bibnamefont
  {Lee}}\ and\ \bibinfo {author} {\bibfnamefont {W.~E.}\ \bibnamefont
  {Pickett}},\ }\href {\doibase 10.1103/PhysRevB.72.174505} {\bibfield
  {journal} {\bibinfo  {journal} {Phys. Rev. B}\ }\textbf {\bibinfo {volume}
  {72}},\ \bibinfo {pages} {174505} (\bibinfo {year} {2005})}\BibitemShut
  {NoStop}%
\bibitem [{\citenamefont {Yuan}\ \emph {et~al.}(2008)\citenamefont {Yuan},
  \citenamefont {Salamon}, \citenamefont {Badica},\ and\ \citenamefont
  {Togano}}]{YUAN20081138}%
  \BibitemOpen
  \bibfield  {author} {\bibinfo {author} {\bibfnamefont {H.~Q}~\bibnamefont
  {Yuan}}, \bibinfo {author} {\bibfnamefont {M.}~\bibnamefont {Salamon}},
  \bibinfo {author} {\bibfnamefont {P.}~\bibnamefont {Badica}}, \ and\ \bibinfo
  {author} {\bibfnamefont {K.}~\bibnamefont {Togano}},\ }\href {\doibase
  https://doi.org/10.1016/j.physb.2007.10.343} {\bibfield  {journal} {\bibinfo
  {journal} {Physica B: Condensed Matter}\ }\textbf {\bibinfo {volume} {403}},\
  \bibinfo {pages} {1138} (\bibinfo {year} {2008})}\BibitemShut {NoStop}%
\bibitem [{\citenamefont {Biswas}\ \emph {et~al.}(2012)\citenamefont {Biswas},
  \citenamefont {Hillier}, \citenamefont {Lees},\ and\ \citenamefont
  {Paul}}]{Biswas2012}%
  \BibitemOpen
  \bibfield  {author} {\bibinfo {author} {\bibfnamefont {P.~K.}\ \bibnamefont
  {Biswas}}, \bibinfo {author} {\bibfnamefont {A.~D.}\ \bibnamefont {Hillier}},
  \bibinfo {author} {\bibfnamefont {M.~R.}\ \bibnamefont {Lees}}, \ and\
  \bibinfo {author} {\bibfnamefont {D.~M.}\ \bibnamefont {Paul}},\ }\href
  {\doibase 10.1103/PhysRevB.85.134505} {\bibfield  {journal} {\bibinfo
  {journal} {Phys. Rev. B}\ }\textbf {\bibinfo {volume} {85}},\ \bibinfo
  {pages} {134505} (\bibinfo {year} {2012})}\BibitemShut {NoStop}%
\bibitem [{\citenamefont {Bauer}\ \emph {et~al.}(2009)\citenamefont {Bauer},
  \citenamefont {Khan}, \citenamefont {Michor}, \citenamefont {Royanian},
  \citenamefont {Grytsiv}, \citenamefont {Melnychenko-Koblyuk}, \citenamefont
  {Rogl}, \citenamefont {Reith}, \citenamefont {Podloucky}, \citenamefont
  {Scheidt}, \citenamefont {Wolf},\ and\ \citenamefont
  {Marsman}}]{Bauer2009PRB}%
  \BibitemOpen
  \bibfield  {author} {\bibinfo {author} {\bibfnamefont {E.}~\bibnamefont
  {Bauer}}, \bibinfo {author} {\bibfnamefont {R.~T.}\ \bibnamefont {Khan}},
  \bibinfo {author} {\bibfnamefont {H.}~\bibnamefont {Michor}}, \bibinfo
  {author} {\bibfnamefont {E.}~\bibnamefont {Royanian}}, \bibinfo {author}
  {\bibfnamefont {A.}~\bibnamefont {Grytsiv}}, \bibinfo {author} {\bibfnamefont
  {N.}~\bibnamefont {Melnychenko-Koblyuk}}, \bibinfo {author} {\bibfnamefont
  {P.}~\bibnamefont {Rogl}}, \bibinfo {author} {\bibfnamefont {D.}~\bibnamefont
  {Reith}}, \bibinfo {author} {\bibfnamefont {R.}~\bibnamefont {Podloucky}},
  \bibinfo {author} {\bibfnamefont {E.-W.}\ \bibnamefont {Scheidt}}, \bibinfo
  {author} {\bibfnamefont {W.}~\bibnamefont {Wolf}}, \ and\ \bibinfo {author}
  {\bibfnamefont {M.}~\bibnamefont {Marsman}},\ }\href {\doibase
  10.1103/PhysRevB.80.064504} {\bibfield  {journal} {\bibinfo  {journal} {Phys.
  Rev. B}\ }\textbf {\bibinfo {volume} {80}},\ \bibinfo {pages} {064504}
  (\bibinfo {year} {2009})}\BibitemShut {NoStop}%
\bibitem [{\citenamefont {Smidman}\ \emph {et~al.}(2014)\citenamefont
  {Smidman}, \citenamefont {Hillier}, \citenamefont {Adroja}, \citenamefont
  {Lees}, \citenamefont {Anand}, \citenamefont {Singh}, \citenamefont {Smith},
  \citenamefont {Paul},\ and\ \citenamefont {Balakrishnan}}]{Smidman2004}%
  \BibitemOpen
  \bibfield  {author} {\bibinfo {author} {\bibfnamefont {M.}~\bibnamefont
  {Smidman}}, \bibinfo {author} {\bibfnamefont {A.~D.}\ \bibnamefont
  {Hillier}}, \bibinfo {author} {\bibfnamefont {D.~T.}\ \bibnamefont {Adroja}},
  \bibinfo {author} {\bibfnamefont {M.~R.}\ \bibnamefont {Lees}}, \bibinfo
  {author} {\bibfnamefont {V.~K.}\ \bibnamefont {Anand}}, \bibinfo {author}
  {\bibfnamefont {R.~P.}\ \bibnamefont {Singh}}, \bibinfo {author}
  {\bibfnamefont {R.~I.}\ \bibnamefont {Smith}}, \bibinfo {author}
  {\bibfnamefont {D.~M.}\ \bibnamefont {Paul}}, \ and\ \bibinfo {author}
  {\bibfnamefont {G.}~\bibnamefont {Balakrishnan}},\ }\href {\doibase
  10.1103/PhysRevB.89.094509} {\bibfield  {journal} {\bibinfo  {journal} {Phys.
  Rev. B}\ }\textbf {\bibinfo {volume} {89}},\ \bibinfo {pages} {094509}
  (\bibinfo {year} {2014})}\BibitemShut {NoStop}%
\bibitem [{\citenamefont {Anand}\ \emph {et~al.}(2011)\citenamefont {Anand},
  \citenamefont {Hillier}, \citenamefont {Adroja}, \citenamefont {Strydom},
  \citenamefont {Michor}, \citenamefont {McEwen},\ and\ \citenamefont
  {Rainford}}]{AnandPhysRevB2011}%
  \BibitemOpen
  \bibfield  {author} {\bibinfo {author} {\bibfnamefont {V.~K.}\ \bibnamefont
  {Anand}}, \bibinfo {author} {\bibfnamefont {A.~D.}\ \bibnamefont {Hillier}},
  \bibinfo {author} {\bibfnamefont {D.~T.}\ \bibnamefont {Adroja}}, \bibinfo
  {author} {\bibfnamefont {A.~M.}\ \bibnamefont {Strydom}}, \bibinfo {author}
  {\bibfnamefont {H.}~\bibnamefont {Michor}}, \bibinfo {author} {\bibfnamefont
  {K.~A.}\ \bibnamefont {McEwen}}, \ and\ \bibinfo {author} {\bibfnamefont
  {B.~D.}\ \bibnamefont {Rainford}},\ }\href {\doibase
  10.1103/PhysRevB.83.064522} {\bibfield  {journal} {\bibinfo  {journal} {Phys.
  Rev. B}\ }\textbf {\bibinfo {volume} {83}},\ \bibinfo {pages} {064522}
  (\bibinfo {year} {2011})}\BibitemShut {NoStop}%
\bibitem [{\citenamefont {Qi}\ \emph {et~al.}(2014)\citenamefont {Qi},
  \citenamefont {Guo}, \citenamefont {Lei}, \citenamefont {Xiao}, \citenamefont
  {Kamiya},\ and\ \citenamefont {Hosono}}]{Qi2014}%
  \BibitemOpen
  \bibfield  {author} {\bibinfo {author} {\bibfnamefont {Y.}~\bibnamefont
  {Qi}}, \bibinfo {author} {\bibfnamefont {J.}~\bibnamefont {Guo}}, \bibinfo
  {author} {\bibfnamefont {H.}~\bibnamefont {Lei}}, \bibinfo {author}
  {\bibfnamefont {Z.}~\bibnamefont {Xiao}}, \bibinfo {author} {\bibfnamefont
  {T.}~\bibnamefont {Kamiya}}, \ and\ \bibinfo {author} {\bibfnamefont
  {H.}~\bibnamefont {Hosono}},\ }\href {\doibase 10.1103/PhysRevB.89.024517}
  {\bibfield  {journal} {\bibinfo  {journal} {Phys. Rev. B}\ }\textbf {\bibinfo
  {volume} {89}},\ \bibinfo {pages} {024517} (\bibinfo {year}
  {2014})}\BibitemShut {NoStop}%
\bibitem [{\citenamefont {Ba\ifmmode \breve{g}\else \u{g}\fi{}c\ifmmode \imath
  \else~\i \fi{}}\ \emph {et~al.}(2019)\citenamefont {Ba\ifmmode \breve{g}\else
  \u{g}\fi{}c\ifmmode \imath \else~\i \fi{}}, \citenamefont {Cin},
  \citenamefont {Uzunok}, \citenamefont {Karaca}, \citenamefont
  {T\"ut\"unc\"u},\ and\ \citenamefont {Srivastava}}]{BaPhysRevB2019}%
  \BibitemOpen
  \bibfield  {author} {\bibinfo {author} {\bibfnamefont {S.}~\bibnamefont
  {Ba\u{g}c\i}},
  \bibinfo {author} {\bibfnamefont {M.}~\bibnamefont {Cin}}, \bibinfo {author}
  {\bibfnamefont {H.~Y.}\ \bibnamefont {Uzunok}}, \bibinfo {author}
  {\bibfnamefont {E.}\ \bibnamefont {Karaca}}, \bibinfo {author}
  {\bibfnamefont {H.~M.}\ \bibnamefont {T\"ut\"unc\"u}}, \ and\ \bibinfo
  {author} {\bibfnamefont {G.~P.}\ \bibnamefont {Srivastava}},\ }\href
  {\doibase 10.1103/PhysRevB.100.184507} {\bibfield  {journal} {\bibinfo
  {journal} {Phys. Rev. B}\ }\textbf {\bibinfo {volume} {100}},\ \bibinfo
  {pages} {184507} (\bibinfo {year} {2019})}\BibitemShut {NoStop}%
\bibitem [{\citenamefont {Keiber}\ \emph {et~al.}(1984)\citenamefont {Keiber},
  \citenamefont {W{\"u}hl}, \citenamefont {Meisner},\ and\ \citenamefont
  {Stewart}}]{keiber1984phonon}%
  \BibitemOpen
  \bibfield  {author} {\bibinfo {author} {\bibfnamefont {H.}~\bibnamefont
  {Keiber}}, \bibinfo {author} {\bibfnamefont {H.}~\bibnamefont {W{\"u}hl}},
  \bibinfo {author} {\bibfnamefont {G.}~\bibnamefont {Meisner}}, \ and\
  \bibinfo {author} {\bibfnamefont {G.}~\bibnamefont {Stewart}},\ }\href
  {https://link.springer.com/article/10.1007/BF00683648} {\bibfield  {journal}
  {\bibinfo  {journal} {Journal of low temperature physics}\ }\textbf {\bibinfo
  {volume} {55}},\ \bibinfo {pages} {11} (\bibinfo {year} {1984})}\BibitemShut
  {NoStop}%
\bibitem [{\citenamefont {Bhattacharyya}\ \emph {et~al.}(2019)\citenamefont
  {Bhattacharyya}, \citenamefont {Panda}, \citenamefont {Adroja}, \citenamefont
  {Kase}, \citenamefont {Biswas}, \citenamefont {Saha}, \citenamefont {Das},
  \citenamefont {Lees},\ and\ \citenamefont {Hillier}}]{Bhattacharyya2019JPCM}%
  \BibitemOpen
  \bibfield  {author} {\bibinfo {author} {\bibfnamefont {A.}~\bibnamefont
  {Bhattacharyya}}, \bibinfo {author} {\bibfnamefont {K.}~\bibnamefont
  {Panda}}, \bibinfo {author} {\bibfnamefont {D.~T.}\ \bibnamefont {Adroja}},
  \bibinfo {author} {\bibfnamefont {N.}~\bibnamefont {Kase}}, \bibinfo {author}
  {\bibfnamefont {P.~K.}\ \bibnamefont {Biswas}}, \bibinfo {author}
  {\bibfnamefont {S.}~\bibnamefont {Saha}}, \bibinfo {author} {\bibfnamefont
  {T.}~\bibnamefont {Das}}, \bibinfo {author} {\bibfnamefont {M.~R.}\
  \bibnamefont {Lees}}, \ and\ \bibinfo {author} {\bibfnamefont {A.~D.}\
  \bibnamefont {Hillier}},\ }\href {\doibase 10.1088/1361-648x/ab549e}
  {\bibfield  {journal} {\bibinfo  {journal} {Journal of Physics: Condensed
  Matter}\ }\textbf {\bibinfo {volume} {32}},\ \bibinfo {pages} {085601}
  (\bibinfo {year} {2019})}\BibitemShut {NoStop}%
\bibitem [{\citenamefont {Panda}\ \emph {et~al.}(2019)\citenamefont {Panda},
  \citenamefont {Bhattacharyya}, \citenamefont {Adroja}, \citenamefont {Kase},
  \citenamefont {Biswas}, \citenamefont {Saha}, \citenamefont {Das},
  \citenamefont {Lees},\ and\ \citenamefont {Hillier}}]{PandaPhysRevB2019}%
  \BibitemOpen
  \bibfield  {author} {\bibinfo {author} {\bibfnamefont {K.}~\bibnamefont
  {Panda}}, \bibinfo {author} {\bibfnamefont {A.}~\bibnamefont
  {Bhattacharyya}}, \bibinfo {author} {\bibfnamefont {D.~T.}\ \bibnamefont
  {Adroja}}, \bibinfo {author} {\bibfnamefont {N.}~\bibnamefont {Kase}},
  \bibinfo {author} {\bibfnamefont {P.~K.}\ \bibnamefont {Biswas}}, \bibinfo
  {author} {\bibfnamefont {S.}~\bibnamefont {Saha}}, \bibinfo {author}
  {\bibfnamefont {T.}~\bibnamefont {Das}}, \bibinfo {author} {\bibfnamefont
  {M.~R.}\ \bibnamefont {Lees}}, \ and\ \bibinfo {author} {\bibfnamefont
  {A.~D.}\ \bibnamefont {Hillier}},\ }\href {\doibase
  10.1103/PhysRevB.99.174513} {\bibfield  {journal} {\bibinfo  {journal} {Phys.
  Rev. B}\ }\textbf {\bibinfo {volume} {99}},\ \bibinfo {pages} {174513}
  (\bibinfo {year} {2019})}\BibitemShut {NoStop}%
\end{thebibliography}

%
\bibliographystyle{apsrev4-1}

\end{document}